\documentclass[longauth]{aa}  

\usepackage{graphicx}
\usepackage[normalem]{ulem}
\usepackage{amsmath,amssymb}
\usepackage{adjustbox} 

\usepackage{epigraph} 
\usepackage{wasysym}
\usepackage{wrapfig}

\usepackage{xcolor}
\usepackage{titlesec} 
\usepackage{changepage} 
\usepackage{txfonts}
\begin{document} 


    \title{FAUST XXVI. The dust opacity spectral indices of protostellar envelopes bridge the gap between interstellar medium and disks.}

    \titlerunning{Continuum emission from the ALMA FAUST Large Program}
    \authorrunning{Cacciapuoti et al.}
   
   \author{Luca Cacciapuoti
          \inst{1}
          \and
          L.Testi 
          \inst{2}
          \and 
          A.J.Maury
          \inst{3,4}
          \and 
          C.J.Chandler
          \inst{5}
          \and 
          N.Sakai 
          \inst{6}
          \and
          C.Ceccarelli
          \inst{7}
          \and 
          C.Codella
          \inst{8}
          \and
          M.De Simone
          \inst{9}
          \and
          L.Podio
          \inst{8}
          \and
          G.Sabatini
          \inst{8}
          \and
          E.Bianchi
          \inst{8}
          \and
          E.Macias
          \inst{9}
          \and
          A.Miotello
          \inst{9}
          \and
          C.Toci
          \inst{9}
          \and
          L.Loinard
          \inst{10,11,12}
          \and
          D.Johnstone
          \inst{13,14}
          \and
          H.B.Liu
          \inst{15,16}
          \and
          Y.Aikawa
          \inst{17}
          \and
          Y.Shirley
          \inst{18}
          \and
          B.Svoboda
          \inst{5}
          \and
          T.Sakai
          \inst{19}
          \and
          T.Hirota
          \inst{6,20}
          \and
          S.Viti
          \inst{21,22}
          \and
          B.Lefloch
          \inst{7}
          \and
          Y.Oya
          \inst{23}
          \and
          S.Ohashi
          \inst{20}
          \and
          S.Feng
          \inst{24}
          \and
          F.Fontani
          \inst{8}
          \and
          Y.Watanabe
          \inst{25}
          \and
          A.Lopez-Sepulcre
          \inst{7,26}
          \and
          Y.Zhang
          \inst{27}
          \and
          C.Vastel
          \inst{28}
          \and
          D.Segura-Cox
          \inst{29}
          \and
          J.Pineda
          \inst{30}
          \and
          A.Isella
          \inst{31}
          \and
          R.S.Klessen
          \inst{32,33}
          \and
          P.Hennebelle 
          \inst{4}
          \and
          S.Molinari
          \inst{34}
          \and
          S.Yamamoto
          \inst{35}
          }

   \institute{European Southern Observatory, Alonso de Cordova 3107, Vitacura, Region Metropolitana de Santiago, Chile
   \and
   Dipartimento di Fisica e Astronomia "Augusto Righi" Viale Berti Pichat 6/2, Bologna
   \and
   Institut d’Estudis Espacials de Catalunya (IEEC), c/Gran Capita, 2-4, E-08034 Barcelona, Catalonia, Spain
   \and
   Universit\'{e} Paris-Saclay, Universit\'{e} Paris Cité, CEA, CNRS, AIM, 91191, Gif-sur-Yvette, France
    \and
    National Radio Astronomy Observatory, PO Box O, Socorro, NM 87801, USA
    \and
    RIKEN Cluster for Pioneering Research, 2-1, Hirosawa, Wako-shi, Saitama 351-0198, Japan
    \and
    Univ. Grenoble Alpes, CNRS, IPAG, 38000 Grenoble, France
    \and
    INAF, Osservatorio Astrofisico di Arcetri, Largo E. Fermi 5, I-50125, Firenze, Italy
    \and
    European Southern Observatory, Karl-Schwarzschild-Strasse 2 D-85748 Garching bei Munchen, Germany
    \and
    Instituto de Radioastronomía y Astrofísica, Universidad Nacional Autonóma de México, Apartado Postal 3-72, Morelia 58090, Michoacán, Mexico
     \and
    Black Hole Initiative at Harvard University, 20 Garden Street, Cambridge, MA 02138, USA
    \and
    David Rockefeller Center for Latin American Studies, Harvard University, 1730 Cambridge Street, Cambridge, MA 02138, USA
     \and
    NRC Herzberg Astronomy and Astrophysics, 5071 West Saanich Rd, Victoria, BC, V9E 2E7, Canada
     \and
    Department of Physics and Astronomy, University of Victoria, Victoria, BC, V8P 5C2, Canada
     \and
    Department of Physics, National Sun Yat-Sen University, No. 70, Lien-Hai Road, Kaohsiung City 80424, Taiwan, R.O.C.         
   \and
     Center of Astronomy and Gravitation, National Taiwan Normal University, Taipei 116, Taiwan
    \and
    Department of Astronomy, The University of Tokyo, 7-3-1 Hongo, Bunkyoku, Tokyo 113-0033, Japan
    \and
    Steward Observatory, 933 N Cherry Avenue, Tucson, AZ 85721, USA
    \and
    Graduate School of Informatics and Engineering, The University of Electro-Communications, Chofu, Tokyo 182-8585, Japan
    \and
    National Astronomical Observatory of Japan, Osawa 2-21-1, Mitakashi, Tokyo 181-8588, Japan
    \and
    Leiden Observatory, Leiden University, PO Box 9513, 2300 RA Leiden, The Netherlands
    \and
    Department of Physics and Astronomy, University College London, Gower Street, London, WC1E 6BT, UK
    \and
    Center for Gravitational Physics, Yukawa Institute for Theoretical Physics, Kyoto University, Oiwake-cho, Kitashirakawa, Sakyo-ku, Kyoto-shi, Kyoto-fu 606- 8502, Japan
    \and
    Department of Astronomy, Xiamen University, Xiamen, Fujian 361005, People’s Republic of China
    \and
    Materials Science and Engineering, College of Engineering, Shibaura Institute of Technology, 3-7-5 Toyosu, Koto-ku, Tokyo 135-8548, Japan
    \and
    Institut de Radioastronomie Millimétrique, 38406 Saint-Martin d’Héres, France
    \and
     Department of Astronomy, Shanghai Jiao Tong University, 800 Dongchuan Road, Minhang, Shanghai 200240, People’s Republic of China
    \and
    IRAP, Université de Toulouse, CNRS, CNES, UPS, Toulouse, France
    \and
    University of Texas at Austin, Department of Astronomy, 2515 Speedway,Stop C1400,Austin, TX78712-1205, USA
    \and
    Max-Planck-Institut für extraterrestrische Physik (MPE), Giessenbachstr. 1, D-85741 Garching, Germany
    \and
    Department of Physics and Astronomy, Rice University, 6100 Main Street, MS-108, Houston, TX 77005, USA
    \and
    Universität Heidelberg, Zentrum für Astronomie, Institut für Theoretische Astrophysik, Albert-Ueberle-Straße 2, 69120 Heidelberg, Germany
    \and
    Universität Heidelberg, Interdisziplinäres Zentrum für Wissenschaftliches Rechnen, Im Neuenheimer Feld 205, 69120 Heidelberg, Germany
    \and
    INAF-Istituto di Astrofisica e Planetologia Spaziali, Via del Fosso del Cavaliere 100, I-00133, Rome, Italy
    \and
    SOKENDAI, Shonan Village, Hayama, Kanagawa 240-0193, Japan
}

   \date{Received March 2025; accepted June 2025}
    
 
  \abstract
   {The (sub-)millimetre dust opacity spectral index ($\beta$) is a critical observable to constrain dust properties, such as the maximum grain size of an observed dust population. It has been widely measured at galactic scales and down to protoplanetary disks. However, because of observational and analytical challenges, quite a gap exists in following the evolution of dust in the interstellar medium (ISM): measures of dust properties across the envelopes that feed newborn protostars and their disks are scarce.}
   {To fill this gap, we use sensitive dust continuum emission data at 1.2 and 3.1 mm from the ALMA FAUST Large Program and constrain the dust opacity submillimetre spectral index around a sample of protostars.}
   {Our high-resolution data, along with a more refined methodology with respect to past efforts, allow us to disentangle disk and envelope contributions in the uv-plane, and thus measure spectral indices for the envelopes uncontaminated by the optically thick emission of the inner disk regions.}
   {Among FAUST sources ($n=13$), we observe a variety of morphologies in continuum emission: compact young disks, extended collapsing envelopes, and dusty outflow cavity walls. First, we find that the young disks in our sample are small (down to $<$9 au) and optically thick. Secondly, we measure the dust opacity spectral index $\beta$ at envelope scales for $n=11$ sources: among these, the $\beta$ of $n=9$ sources had never been constrained in the literature. We effectively double the number of sources for which the dust opacity spectral index $\beta$ has been measured at these scales.
   Third, combining the available literature measurements with our own (adding up to a total $n=18$), we show how envelope spectral indices distribute between ISM-like and disk-like values, bridging the gap in the inferred dust evolution. Finally, we statistically confirm a significant correlation between $\beta$ and the mass of protostellar envelopes, previously suggested in the literature.}
   {Our findings indicate that the dust optical properties smoothly vary from the ISM ($\gg$ 0.1 parsec), through envelopes ($\sim$ 500-2000 au) and all the way down to protoplanetary disks ($<$ 200 au). Multi-wavelength surveys including longer wavelengths and in controlled star-forming regions are needed to further this study and make more general claims on dust evolution in its pathway from cloud to disks.}
   
   \keywords{ISM: dust, extinction, planets and satellites: formation, protoplanetary disks, techniques: interferometric, submillimeter: planetary systems}

\maketitle
%

\section{Introduction}
\label{sec:intro}

Even accounting for approximately a percent of the material that permeates the interstellar medium (ISM), dust grains play several crucial roles in star and planet formation (\citealt{Tielens2010}, \citealt{Draine2011}), and thus their evolution needs to be precisely constrained across space and time in star-forming regions. 
\begin{table*}[h]
    \centering
    \begin{tabular}{l|c|c|c|c|c|c|r}
    \hline \hline 
        Source Name & L$_{\textrm{bol}}$ (L$_{\odot}$) & T$_{\textrm{bol}}$ (K) & M$_{\textrm{env}}$ (M$_{\odot}$)& inc (deg) & Binary? & Class & Region (distance) \\
        \hline  
        IRAS 15398-3359 &  0.92$^{(a)}$ & 44$^{(a)}$ & 1.2$^{(a)}$ & 70$^{(s)}$ & N & 0 & Lupus (155 pc) \\
        
        CB68 & 0.86$^{(b)}$ & 41$^{(c)}$ & 0.9$^{(b)}$ & 70$^{(t)}$ & N & 0 & Isolated (137 pc) \\
        
        L483 & 9.0$^{(d)}$ &  50$^{(a)}$ & 4.4$^{(d)}$ & 83$^{(u)}$ & N& 0 & Aquila (200 pc) \\
        
        Elias 29 & 13.0$^{(e)}$ & 391$^{(f)}$& 0.47$^{(e)}$ & 60$^{(v)}$ & N& I & Ophiuchus (137 pc) \\
        
        VLA1623A & 2.6$^{(g)}$ & 55$^{(h)}$ & 0.8$^{(h)}$ & 55$^{(w)}$ & Y & 0 & Ophiuchus (137 pc)\\
        
        GSS30 & 1.7$^{(i)}$ & 50$^{(i)}$ & 0.098$^{(e)}$ & 64$^{(y)}$ & N &I & Ophiuchus (137 pc) \\
        
        RCra IRS7B & 5.1$^{(i)}$ & 88$^{(i)}$ & 6.3$^{(l)}$ & 65$^{(z)}$ & Y & 0/I & Corona Australis (130 pc)\\
        
        BHB07-11 & 2.2$^{(m)}$ & 65$^{(n)}$ & 0.090$^{(m)}$ & 85$^{(m)}$ & Y & 0/I &  Pipe (145 pc) \\        
        
        IRS63 & 1.3$^{(i)}$ &  348$^{(i)}$ & 0.096$^{(e)}$ & 47$^{(i)}$ & N &I &  Ophiuchus (137 pc)\\
        
        NGC1333 IRAS4A & 9.1$^{(o)}$ & 29$^{(p)}$ & 12.2$^{(p)}$ & 65$^{(x)}$ & Y & 0  & Perseus (235 pc) \\
        
        NGC1333 IRAS4C & 1.1$^{(q)}$ & 31$^{(q)}$ & 1.2$^{(q)}$ & 75$^{(xy)}$ & N& 0 & Perseus (235 pc)\\

        L1527 IRS$^*$ & 1.3$^{(i)}$ & 41$^{(i)}$ & 1.2$^{(r)}$ & 85$^{(a)}$ & N &0/I & Taurus (137 pc)\\

        L1551-IRS5 & 24.5 & 92$^{(g)}$ & 1.6$^{(g)}$ & 60$^{(xz)}$ & Y& I & Taurus (147 pc)\\

        \hline 
    \end{tabular}
    \caption{Main properties of the ALMA FAUST Large Program sources analysed in this work. \textit{References:} $^{(a)}$ \citealt{Jorgensen2013}; $^{(b)}$ \citealt{Valle2000}; $^{(c)}$ \citealt{Launhardt2013}; $^{(d)}$ \citealt{Jorgensen2002}; $^{(e)}$ \citealt{Jorgensen2009}; $^{(f)}$ \citealt{Chen1995}; $^{(g)}$ \citealt{Froebrich2005}; $^{(h)}$ \citealt{Myers1998}; $^{(i)}$ \citealt{Ohashi2023}; $^{(l)}$ \citealt{vanKempen2009}; $^{(m)}$ \citealt{Alves2017}; $^{(n)}$ \citealt{Evans2023}; $^{(o)}$ \citealt{Andre2010}; $^{(p)}$ \citealt{Sadavoy2014}; $^{(q)}$ \citealt{Mottram2017}; $^{(r)}$ \citealt{Motte2001}; $^{(s)}$ \citealt{okoda2018}; $^{(t)}$ \citealt{Imai2022}; $^{(u)}$ \citealt{Oya2018}; $^{(v)}$ \citealt{Miotello2014}; $^{(w)}$ \citealt{Sadavoy2019}; $^{(y)}$ \citealt{Santamaria-Miranda2024} ; $^{(z)}$ \citealt{Takakuwa2024}; $^{(x)}$ \citealt{Yildiz2012}; $^{(xy)}$ \citealt{Segura-Cox2016}; $^{(xz)}$ \citealt{Cruz-SaenzdeMiera2019}. Note: (a) L1527 IRS was analysed in the pilot study of \citealt{Cacciapuoti2023}.}
    \label{tab:faust_sample}
\end{table*}

From the slope of the submillimetre SED ($\alpha$), the dust absorption opacity $\kappa \propto \nu^{\beta}$ spectral index in the same wavelength range ($\beta$) can be derived. The latter carries important information on cosmic dust: among others, $\beta$ anti-correlates with the maximum grain size of the distribution (e.g. \citealt{Natta2007}, \citealt{Testi2014}, \citealt{Ysard2019}). According to these studies, values of $\beta \gtrsim 1.5$ are consistent with dust grains smaller than about 10 $\mu$m, while grain sizes larger than $\sim$100 $\mu$m flatten the dust opacity and $\beta \lesssim 1$.

In the ISM, the submillimetre dust opacity spectral index has largely been characterised to be $\beta \sim$ 1.5-1.9 (e.g., \citealt{Schwartz1982}, \citealt{Planck2014}, \citealt{Juvela2015}), constraining the dust in the diffuse medium to be (sub-)micron sized. Recent observations in different wavelength ranges have been challenging models of dust coagulation in the ISM and in denser cores. Dust models often assume the ISM dust to have an MRN-like distribution \citep{Mathis1977} everywhere, and thus with sizes in the [0.005, 1] $\mu$m range. Near-infrared observations have, on the contrary, hinted at the presence of larger grains (up to 5$\mu$m) in denser regions of the ISM via extinction measurements \citep{Roy2013}, scattered light ``coreshine'' effect  (\citealt{Pagani2010}, \citealt{Steinacker2010}, \citealt{Steinacker2015}), and even recent JWST spectroscopy \citep{Dartois2024}. Several constraints thus point in the direction of dust evolution from the very diffuse ISM to its denser cores, in which star formation begins.

On the smaller end of physical scales, large surveys to measure submillimetre spectral indices in protoplanetary disks have been carried out in order to pinpoint the onset of planet formation, understood as the growth of dust into pebbles that will form planetesimals in disks. These surveys yielded $\alpha < 3$ (and $\beta < 1$) for the majority of disks (e.g., \citealt{Ricci2010}, \citealt{Tazzari2021}, \citealt{Chung2024}, \citealt{Garufi2025}). Based on such studies, important constraints were derived on the evolution of dust in planet-forming environments, and also indirect evidence was found for protoplanetary disk substructures - suggested by the low $\alpha$/$\beta$, hinting to the high optical depth of localised regions of disks (e.g., \citealt{Ricci2012}, \citealt{Delussu2024}). Beyond total intensity measurements, polarization studies also point to dust growth in disks (\citealt{Zhang2023}, \citealt{Lin2024}). It is today well understood that dust grains grow in protoplanetary disks substructures at least to cm sizes, before possibly quickly building up km-sized planetesimals via mechanisms such as streaming instabilities (\citealt{Youdin2005}, \citealt{Scardoni2024}).

However, a gap is still open in our knowledge of the life cycle of dust from very diffuse ISM to very dense disks. Surveys aiming to characterize the spectral indices of environments between the ISM cores and protoplanetary disks, i.e. protostellar envelopes, are lacking. This is mainly because of two observational difficulties.
First, extinction-based studies in the optical-, and near-/mid-infrared can only be used to characterize dust up to visual extinctions of a few magnitudes, after which the higher densities do not allow any background photon to pierce the region and reach our detectors. Moreover, in order to study much more compact and denser regions than the ones considered in the above mentioned studies, the high spatial ($\sim$ 1'') resolution probed by (sub-)millimetre interferometers becomes a necessity. However, thermal dust emission at these frequencies becomes fainter and fainter away from a radiation source. Thus, the emission of protostellar envelopes is much fainter than the one of protoplanetary disks, making it harder to survey in a large number of objects.
Due to the time-consuming observations, which need to be characterised by high resolution, large maximum recoverable scales and high sensitivities, collapsing protostellar envelopes have never been systematically surveyed to constrain their dust properties, and yet these are central to understand many aspects of star and planet formation.

The evolution of dust grains in protostellar envelopes can affect a set of conditions for star and planet formation. The infalling dust grains represent the initial distribution of solids input in the planet-forming disks. Moreover, dust grains are charged and thus coupled with the protostar's magnetic fields: this leads to a ``magnetic braking effect'' which slows down collapse and regulates the sizes of rotationally-supported protoplanetary disks (\citealt{Hennebelle2009}, \citealt{Zhao2016}, \citealt{Vallucci-Goy2024}). Additionally, if grains are large in envelopes, they might decouple from and infall faster than the gas, leading to increases in the dust-to-gas ratio in the disks, critical to form early planetesimals (\citealt{Youdin2005}, \citealt{Cridland2022}). 
Finally, dust grain properties are also invoked to explain the observed thermal polarised emission in the ISM via radiative alignment torques theory (\citealt{Lazarian2007}, \citealt{Reissl2020}), mechanical alignment along gas flows (\citealt{Hall1949}, \citealt{Reissl2023}) and radiation alignment (\citealt{Andersson2015}). As these mechanisms depend on grain properties, refined constraints on dust grains in envelopes is precious to model and explain polarisation observations.

A few works have attempted to perform such measurements and have found ambiguous results. \citet{Kwon09} and \citet{Miotello2014} measured low $\beta$ values ($\leq 1$) in five objects, but with large relative uncertainties ($> 50$\%) due to the relatively poor sensitivity of the observations, the need to combine data from different facilities and the large calibration uncertainties of the (mix of) instruments at play. Nonetheless, they interpreted such results as possible hints of grain growth in collapsing envelopes for the first time. Using similar methods, other works have not found indications of such an evolution, and thus first hinted at possible tensions as to whether growth of dust is a viable process at envelope scales or not (\citealt{Agurto-Gangas2019}, \citealt{Cacciapuoti2024b}). To date, the only rather extended study by \citet{Galametz2019} focused on the Plateau de Bure Interferometer (PdBI) CALYPSO sample of Class 0/I sources presented by \citet{Maury2018}. For nine (out of sixteen) sources for which they could measure $\beta$, they found examples of both relatively small and large $\beta$ values (in the range 0.5$-$1.4), with a variety of gradients from the outer to inner envelope that would suggest the presence of submillimetre grains in a few envelopes.

\begin{figure*}
    \centering
    \includegraphics[width=\linewidth] {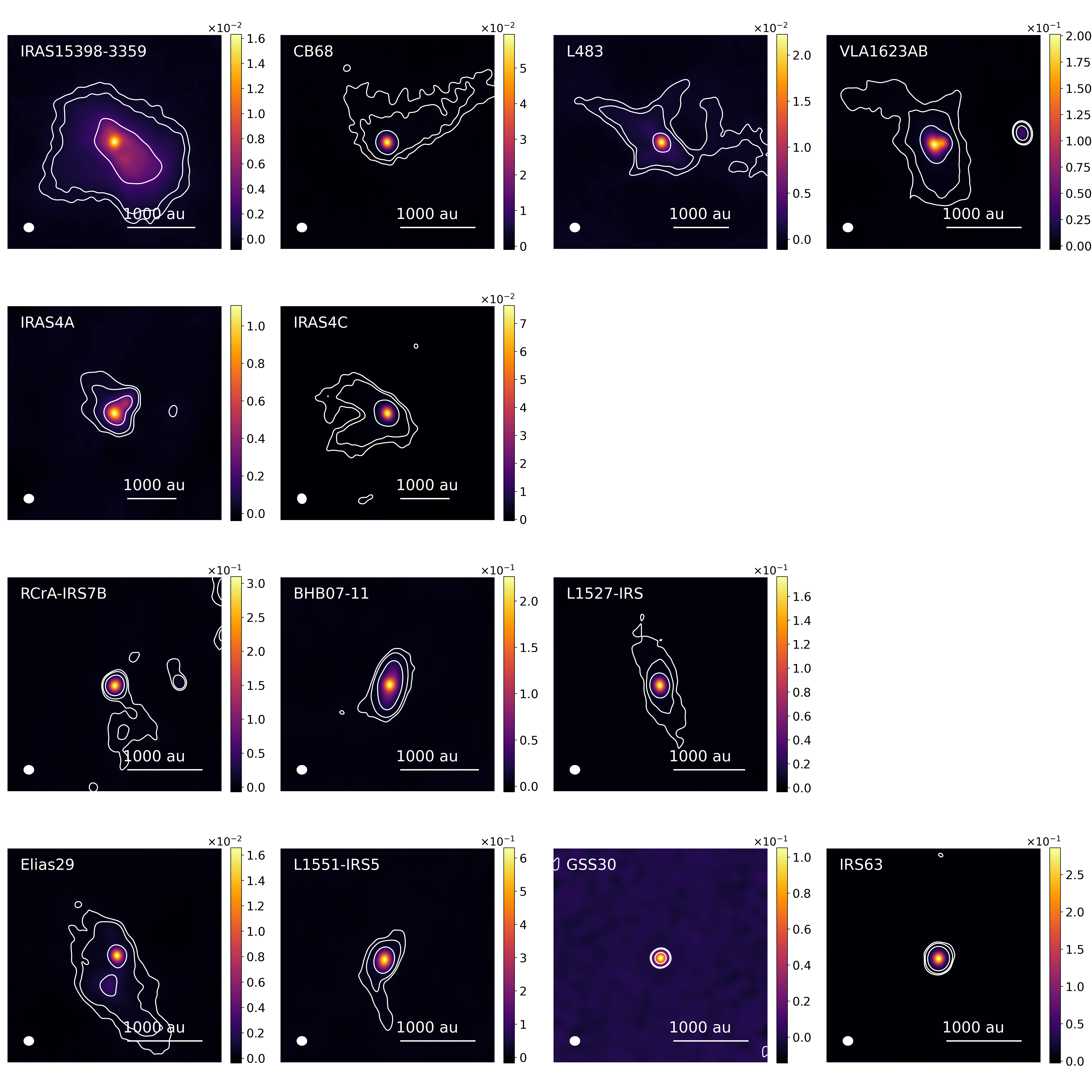}
    \caption{The dust continuum emission of the ALMA FAUST large program sample at 1.2 mm, colorbar units in Jy/beam. The synthesised beam is depicted in the lower left as a white ellipse and reported, along with the rms of each map, in Tab. \ref{tab:map_stats}. The white contours are at [5, 10, 50]$\sigma$ levels. The panels are centred on the source and cover 20 arcsec across. The first two rows are Class 0 sources, the third row is for Class 0/I sources, and the last row is for Class I sources (Tab. \ref{tab:faust_sample}).}
    \label{fig:sample_1.2 mm}
\end{figure*}

On the theoretical side, simulations of dust evolution during collapse have a hard time reproducing such observations (e.g., \citealt{Ormel2009}, \citealt{Bate2022}, \citealt{Lebreuilly2023a}), and conclude that grains can grow in envelopes only up to $\sim$2 $\mu$m. Whether we are missing some fundamental aspect on how dust evolves during collapse, or if there are any observational biases against the few measured low $\beta$ values, is a question that requires improving the number of sources with robust measurements (e.g., \citealt{Birnstiel2024}, \citealt{Perotti2024}). 
Theoretical models are now longing for tighter constraints on the dust distribution of these early, large-scale environments to gauge their effect on systems' evolution. 

In this paper, thanks to the ALMA FAUST Large Program, we use sensitive and high-resolution (50 au) observations that can help us overcome the limitation of previous attempts. We measure in a robust and consistent way the dust opacity spectral index in the envelope of 11 Class 0/I sources (of which 9 independent and 2 overlapping with the CALYPSO sample), with the aim of exploring the gap between ISM and disk's spectral index values. 
In Section \ref{obs_faust}, we present the sample and observations. In Section \ref{methods_faust}, we present the method with which we disentangle the disk and envelope emission. In Section \ref{results_alpha_beta}, we present the measurements of spectral indices for each source. We discuss our findings in Section \ref{discuss_faust} and sum up our conclusions and prospects in Section \ref{conclusions_faust}.

\section{Observations and sample}
\label{obs_faust}
 \begin{figure*}[t]
    \centering
    \includegraphics[width=0.85\linewidth] {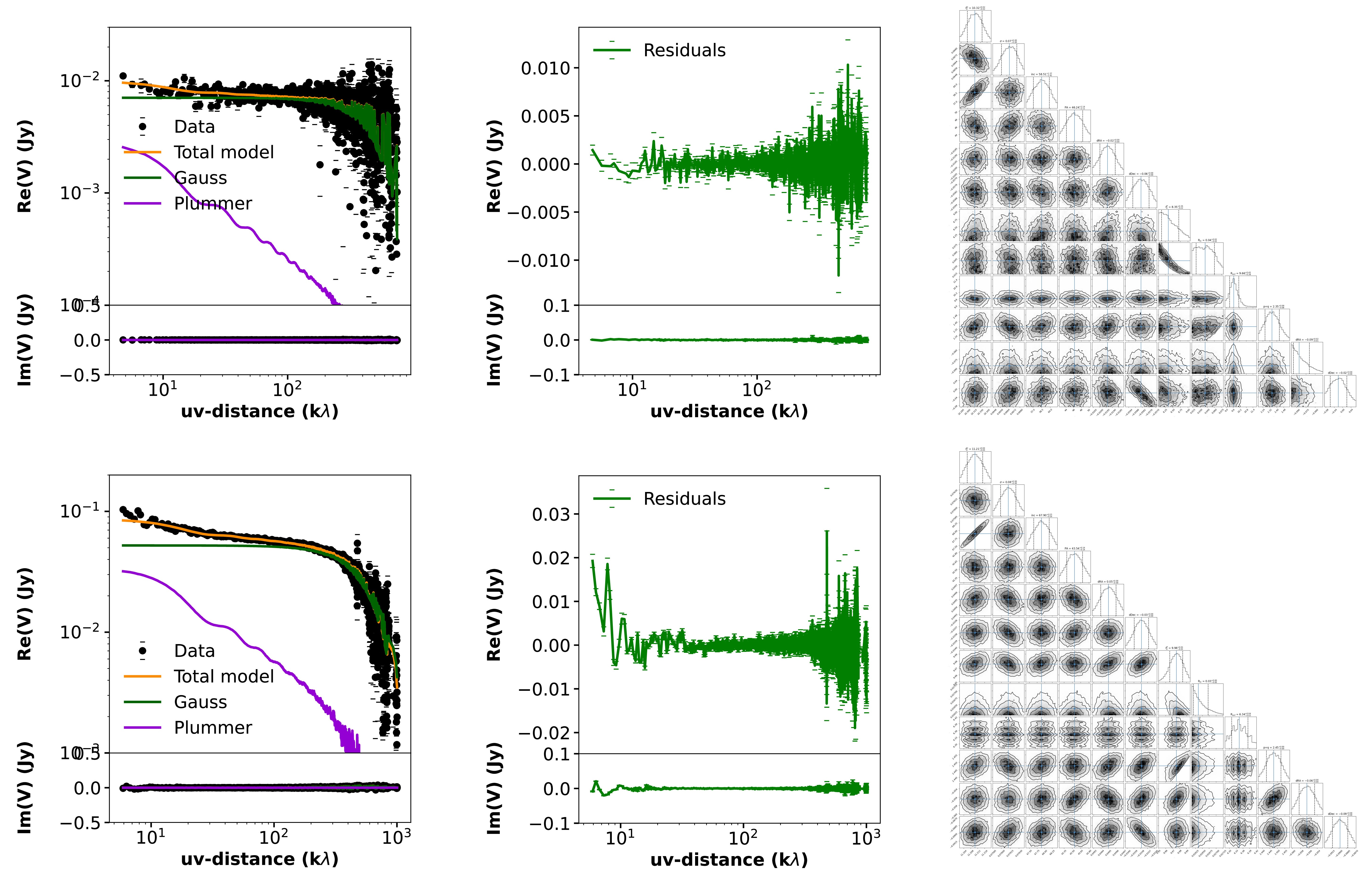}
    \caption{Plummer plus Gaussian best fit (orange) is overplotted on the real and imaginary parts of the visibilities for the B3 (upper panel, black points) and B6 (lower panel) observations of CB68. 
    The Plummer only (violet line) and Gaussian only (green line) components of the total model are also shown. The wiggles in the model are due to its sampling on the uv-points of the observations. The residuals of the model are shown in green at the side of each plot. The corner plots are shown for completeness and the parameters values reported in Table \ref{tab:fits_results} for readability.} 
\label{fig:cb68_fit}
\end{figure*}
In this work, we make use of data from the ALMA Large Program ``Fifty AU Study of the chemistry in the disk/envelope system of Solar-like protostars'' (FAUST, project-ID 2018.1.01205.L, PI: Yamamoto, S.). The main goal of FAUST is to reveal and quantify the variety in the chemical composition of the envelope/disk system of Solar-like Class 0 and I protostars (see \citealt{Codella2021} for further details and also \citealt{faustiii}, \citealt{faustiv}, \citealt{faustvi}, \citealt{faustviii}, \citealt{faustxii}, \citealt{faustxvii} for a gallery of results). In that context, dust absorption opacity $\kappa$ is critical when studying line emission, and the only way to properly take it into account in chemistry studies is to spatially resolve the continuum emission at two or more frequencies (e.g., \citealt{DeSimone2020}, \citealt{Frediani2025}). Moreover, dust grain size distributions are important from a chemical perspective as grains act as facilitators for important reactions in the ISM, even of the simplest molecules \citep{Ceccarelli2022}.

For this reason also, the FAUST program includes observations of young sources in two wavelength windows within the 1.2 mm (239 GHz) and 3.1 mm (95 GHz) atmospheric windows. Moreover, FAUST observations cover a range of baselines that ensure a resolution of 0$\farcs$3-0$\farcs$4 and a maximum recoverable scale of up to 14$\farcs$4 which, given typical distances of the sources (150-300 pc), yields observations that are sensitive to physical scales in the range 50-3000 au, thus allowing proper characterisation of both disks and envelopes. 

A summary of the main properties for the FAUST target sources considered in this study is presented in Table \ref{tab:faust_sample}. Among these, we note that L1527 IRS was already analysed in the pilot study of \citet{Cacciapuoti2023}.

\subsection{Data calibration, reduction and imaging}

The FAUST data were calibrated using a modified version of the ALMA pipeline, using the Common Astronomical Software Applications (CASA, \citealt{CasaTeam2022}), version 5.6.1-8. This included a correction for errors introduced by the per-channel normalisation of data by the ALMA correlator\footnote{https://help.almascience.org/kb/articles/what-errors-could-originate-from-the-correlator-spectral-normalisation-and-tsys-calibration}. Line-free Local Standard of Rest Kinematic (LSRK) frequency ranges were identified by visual inspection and averaged per spectral window, and initial continuum images were produced for each separate ALMA configuration. These were then used as initial models for subsequent per-configuration, phase-only self-calibration followed by amplitude and phase self-calibration. Great care was taken to ensure that the models were as complete as possible to avoid changing the overall flux density scale of the data when doing amplitude self-calibration. The sources are sufficiently bright that ``per-integration'', phase-only self-calibration was possible. For the amplitude self-calibration, the approach was ``per-scan''. The per-configuration datasets were then aligned across configurations in both phase and amplitude, again using a self-calibration technique. Improvements in the dynamic range of more than an order of magnitude for the final images were often achieved using this technique for all setups. The inclusion of absolute flux calibration uncertainties is central when computing spectral indices, since they depend on the flux values at each wavelength. Within FAUST, the default calibrators were used for the flux calibration, and thus it is appropriate to use the flux calibration uncertainties as suggested in the ALMA Techincal Handbook (see also Section \ref{results_alpha_beta}). More details on the FAUST data calibration and reduction can be found in Chandler et al., (in prep.).

In this work, we imaged the continuum data products using a Briggs weighting scheme with a robust parameter of 1 and restraining the utilized uv-distances in the range 6-150 k$\lambda$, in order to highlight the contribution of the large scale emission in which we are interested. We applied a common restoring beam of 1 $\times$ 1 arcsec. We show the primary--beam--corrected images in Fig. \ref{fig:sample_1.2 mm} and Fig. \ref{fig:sample_3.1 mm} in Appendix \ref{app:images}. The synthesised beam of each map and its noise (rms) have been reported in Tab. \ref{tab:map_stats}.

These images were mainly useful to inspect the morphology of the source and visually determine which model to be fit in the uv-plane (see section \ref{methods_faust}). However, for the sources for which we attempt to measure $\alpha$ in the image plane (IRAS4A, BHB07-11, VLA1623A, and IRAS15398-3359; see section \ref{sec:iras4a} and appendix \ref{app:2dmaps}), we generate maps with matching synthesised beams using the \texttt{uvtaper} parameter of the CASA \texttt{tclean} routine. We note that in the uv-plane analysis we compute $\alpha$ across azimuthally averaged baseline bins common between 1.2 and 3.1 mm, so that the bin fluxes used for the computation are recovered from the same physical scales on the sky.

\section{Modeling in the $uv$-plane}
\label{methods_faust}

As we are interested in measuring the dust opacity spectral index and its gradient across protostellar envelopes while correcting for the disk contamination, a multi-scale approach needs to be taken. While very high ($<0\farcs1$) resolution images across observing wavelengths of protoplanetary disks have allowed the community to reliably measure spectral indices in the image plane (e.g., \citealt{CarrascoGonzalez19}, \citealt{Sierra2020}, \citealt{Macias2021}), the need for a multi-scale (50-2000 au) study to trace dust evolution in envelopes calls for a different approach. Indeed, generating an image from interferometric visibilities is a modeling process in which a certain weighting scheme is associated to the visibilities to be imaged. Thus, the final model image will emphasize more compact structures or more extended emission based on which baselines were weighted more among the available ones. For this reason, making an image that ideally traces the source at high resolution and recovers the large-scale emission is challenging. In other words, it is hard to spatially disentangle the contribution of compact and extended structures in the image plane and thus evaluate the contamination of one onto the other when measuring intensity-dependent quantities, like spectral indices. An additional uncertainty in computing spectral index maps naturally arises from sparse uv-coverage, since its variable density also affects the reconstructed sky maps. 

\begin{table*}[h!]
\begin{center}
\renewcommand{\arraystretch}{2}
\begin{adjustbox}{width=\textwidth}
\begin{tabular}{ccccccccccc}
\hline \hline
Source & Band & Gauss f0 (Jy/sr) & Gauss $\sigma$ ('') & Gauss inc (deg) & Gauss PA (deg) & Plummer f0 (Jy/sr) & Plummer \textrm{R$_{\textrm{i}}$} & Plummer \textrm{R$_{\textrm{out}}$} & Plummer \textrm{p+q} & $\chi^2$ \\  \hline 
IRAS1538-3359 & 3 & 10.22$^{+0.13}_{-0.19}$ & 0.03$^{-0.001}_{+0.001}$ & 44$^{-3}_{+4}$ & -73$^{-4}_{+8}$ & 8.02$^{+0.02}_{-0.01}$ & 0.1$^{+0.01}_{-0.01}$ & 8$^{+0.5}_{-0.3}$ & 1.84$^{+0.01}_{-0.01}$ & 1.0\\
-& 6 & 10.95$^{+0.02}_{-0.01}$ & 0.03$^{-0.001}_{+0.001}$ & 41$^{-1}_{+1}$ & -79$^{-1}_{+1}$ & 8.65$^{+0.01}_{-0.01}$ & 0.5$^{+0.02}_{-0.01}$ & 7.9$^{+0.1}_{-0.3}$ & 1.92$^{+0.02}_{-0.01}$ & 1.2\\  \hline
CB68   & 3& 10.32$^{+0.02}_{-0.02}$ & 0.07$^{+0.01}_{-0.01}$ & 58.5$^{+1.3}_{-1.4}$& 46.2$^{+2}_{-2}$& 8.35$^{+0.35}_{-0.22}$& 0.04$^{+0.02}_{-0.02}$& 9.7$^{+0.4}_{-0.4}$ & 2.35$^{+0.08}_{-0.09}$ & 1.0\\ 
-& 6& 11.21$^{+0.01}_{-0.03}$ & 0.08$^{+0.01}_{-0.01}$ & 67.9$^{+0.25}_{-0.25}$& 43.5$^{+0.1}_{-0.1}$ & 9.98$^{+0.01}_{-0.01}$& 0.03$^{+0.001}_{-0.001}$& 6.34$^{+0.03}_{-0.04}$ & 2.45$^{+0.01}_{-0.01}$ & 1.1\\ \hline
L483 & 3& 9.61$^{+0.01}_{-0.01}$& 0.07$^{+0.01}_{-0.01}$& 31.9$^{+1.3}_{-1.4}$& -37$^{+0.1}_{-0.04}$& 8.13$^{+0.1}_{-0.07}$& 0.02$^{+0.001}_{-0.001}$& 13.9$^{+0.1}_{-0.5}$& 1.92$^{+0.02}_{-0.02}$ & 1.1\\ 
-& 6& 10.54$^{+0.01}_{-0.01}$& 0.08$^{+0.02}_{-0.01}$ & 50$^{+0.1}_{-0.1}$ & -38$^{+0.3}_{-0.3}$& 8.22$^{+0.01}_{-0.01}$& 0.5$^{+0.01}_{-0.01}$& 6.75$^{+0.01}_{-0.01}$& 2.02$^{+0.01}_{-0.01}$& 1.1\\ \hline
Elias 29  & 3& 10.61$^{+0.11}_{-0.09}$ & 0.03$^{+0.003}_{-0.001}$& 50$^{+14}_{-15}$& 15$^{+13}_{-14}$ & 8.09$^{+0.09}_{-0.06}$ & 0.02$^{+0.001}_{-0.001}$& 6.11$^{+0.14}_{-0.13}$& 1.88$^{+0.04}_{-0.04}$ & 1.3 \\ 
-& 6& 10.70$^{+0.01}_{-0.01}$& 0.05$^{+0.005}_{-0.002}$& 40$^{+0.01}_{-0.03}$& -30$^{+0.02}_{-0.01}$& 8.55$^{+0.01}_{-0.01}$& 0.05$^{+0.001}_{-0.003}$& 5.40$^{+0.004}_{-0.01}$& 1.72$^{+0.05}_{-0.04}$ & 1.0 \\ \hline
VLA1623A  & 3& 9.70$^{+0.01}_{-0.01}$ & 0.37$^{+0.003}_{-0.002}$ & 76.8$^{+0.2}_{-0.2}$& 90$^{+0.00}_{-0.01}$& 10.49$^{+0.03}_{-0.03}$& 0.01$^{+0.001}_{-0.001}$&5.97$^{+0.03}_{-0.03}$ & 2.38$^{+0.01}_{-0.01}$ & 1.1\\ 
-& 6& 10.54$^{+0.001}_{-0.001}$& 0.34$^{+0.002}_{-0.002}$& 76.2$^{+0.03}_{-0.04}$ & 90$^{+0.0}_{-0.01}$ & 10.59$^{+0.002}_{-0.003}$& 0.13$^{+0.01}_{-0.01}$ & 7.9$^{+0.1}_{-0.1}$ & 2.73$^{+0.02}_{-0.01}$ & 2.0 \\ \hline
IRAS4A1  & 3& 10.17$^{+0.01}_{-0.01}$ & 0.12$^{+0.001}_{-0.002}$ & 38.5$^{+0.2}_{-0.2}$ & 20$^{+0.1}_{-0.2}$ & 10.54$^{+0.01}_{-0.02}$ & 0.13$^{+0.02}_{-0.03}$ & 3.31$^{+0.04}_{-0.03}$ & 2.98$^{+0.02}_{-0.02}$&  25 \\ 
-& 6& 11.18$^{+0.01}_{-0.01}$ & 0.14$^{+0.01}_{-0.02}$ & 45$^{+1}_{-1}$ & -66$^{+1}_{-2}$ & 10.93$^{+0.01}_{-0.03}$ & 0.20$^{+0.02}_{-0.02}$ & 2.94$^{+0.01}_{-0.02}$ & 2.53$^{+0.01}_{-0.02}$ &  35 \\ \hline
IRAS4A2      & 3& - & - & - & - & 10.08$^{+0.02}_{-0.001}$ & 0.07$^{+0.001}_{-0.006}$ & 3.88$^{+0.1}_{-0.3}$ & 2.48$^{+0.01}_{-0.01}$ &  25 \\ 
-& 6& - & - & - & - & 10.89$^{+0.01}_{-0.02}$ & 0.05$^{+0.001}_{-0.002}$ & 8$^{+0.5}_{-0.4}$ & 2.30$^{+0.02}_{-0.02}$ &  35 \\ \hline
L1551-IRS5      & 3& 10.84$^{+0.01}_{-0.01}$ & 0.06$^{+0.001}_{-0.001}$ & 41$^{+0.1}_{-0.2}$ & -20$^{+0.2}_{-0.1}$ & 10.46$^{+0.02}_{-0.01}$ & 0.07$^{+0.001}_{-0.002}$ & 14$^{+0.02}_{-0.10}$ & 3.18$^{+0.02}_{-0.02}$ &  11 \\ 
-& 6& 11.61$^{+0.002}_{-0.001}$ & 0.09$^{+0.001}_{-0.001}$ & 55$^{+0.3}_{-0.1}$ & -29$^{+0.05}_{-0.1}$ & 11.05$^{+0.001}_{-0.001}$ & 0.12$^{+0.001}_{-0.001}$ & 9.97$^{+0.09}_{-0.05}$ & 2.85$^{+0.02}_{-0.02}$ &  12 \\ \hline 
IRAS4C\Large{\textcolor{red}{$^{*}$}}      & 3& 10.00$^{+0.004}_{-0.004}$& 0.13$^{+0.02}_{-0.02}$ & 62.3$^{+0.2}_{-0.2}$ & 14.2$^{+0.2}_{-0.2}$ & -&- &- &- & 13  \\ 
-& 6&  10.38$^{+0.05}_{-0.05}$ & 0.15$^{+0.02}_{-0.02}$ & 70.2$^{+1.4}_{-1.3}$ & 16.3$^{+0.3}_{-0.3}$ & 10.36$^{+0.04}_{-0.04}$ & 0.07$^{+0.004}_{-0.01}$ & 5.7$^{+0.02}_{-0.02}$ & 2.28$^{+0.01}_{-0.01}$ & 12  \\ \hline \hline
Source & Band & Gauss f0 (Jy/sr) & Gauss $\sigma$ ('') & Gauss inc (deg) & Gauss PA (deg) & $\delta$ f0 (Jy/sr) & - & - & -& $\chi^2$\\  \hline 
GSS30\Large{\textcolor{red}{$^{*}$}}       & 3& 10.07$^{+0.01}_{-0.01}$ & 0.23$^{+0.005}_{-0.005}$& 68.7$^{+0.1}_{-0.09}$& -70$^{+0.11}_{-0.11}$& 14.58$^{+0.01}_{-0.01}$ & - & - & - & 1.1 \\ 
-& 6& 10.34$^{+0.02}_{-0.01}$& 0.28$^{+0.01}_{-0.02}$& 62$^{+0.12}_{-0.12}$& -69$^{+0.13}_{-0.15}$&  14.84$^{+0.01}_{-0.01}$& -& -& - &  13 \\ \hline
IRS63\Large{\textcolor{red}{$^{*}$}}       & 3& 9.79$^{+0.01}_{-0.01}$ & 0.22$^{+0.003}_{-0.004}$ & 49$^{+0.1}_{-0.12}$ & -27.4$^{+0.16}_{-0.17}$ & 15.02 & - & - & - & 1.5 \\ 
-& 6& 10.72$^{+0.01}_{-0.01}$ & 0.24$^{+0.002}_{-0.002}$& 47$^{+0.1}_{-0.12}$ & -28$^{+0.3}_{-0.2}$ & 15.57 & -&- &- & 12 \\ \hline \hline              
Source & Band & Gauss$_1$ f0 (Jy/sr) & Gauss$_1$ $\sigma$ ('') & Gauss$_1$ inc (deg) & Gauss$_1$ PA (deg) & Gauss$_2$ f0 (Jy/sr) & Gauss$_2$ $\sigma$ ('') & Gauss$_2$ inc (deg) & Gauss$_2$ PA (deg) & $\chi^2$\\  \hline 
BHB07-11  & 3& 9.08$^{+0.02}_{-0.02}$& 0.32$^{+0.01}_{-0.02}$& 53$^{+2}_{-2}$& 89$^{+1}_{-1}$& 8.07$^{+0.11}_{-0.05}$& 1.62$^{+0.02}_{-0.01}$& 52$^{+8}_{-4}$& -24$^{+5}_{-3}$ & 1.3 \\ 
-& 6& 10.37$^{+0.001}_{-0.002}$ & 0.25$^{+0.02}_{-0.02}$& 40.3$^{+0.1}_{-0.1}$& -60$^{+0.2}_{-0.2}$& 9.31$^{+0.002}_{-0.001}$& 1.59$^{+0.01}_{-0.01}$& 53.0$^{+0.2}_{-0.2}$& -21.6$^{+0.2}_{-0.1}$& 1.2 \\ \hline
RCra IRS7B  & 3& 10.55$^{+0.02}_{-0.0}$ & 0.16$^{+0.01}_{-0.02}$ & 67.7$^{+0.2}_{-0.2}$ & -64.3$^{+0.4}_{-0.4}$  & 10.18$^{+0.01}_{-0.01}$ & 0.07$^{+0.001}_{-0.001}$ & 62.3$^{+0.7}_{-0.7}$ & -75$^{+0.85}_{-0.85}$ & 15  \\ 
-& 6& 11.28$^{+0.003}_{-0.002}$ & 0.18$^{+0.01}_{-0.01}$ & 68.6$^{+0.03}_{-0.01}$ & -64.8$^{+0.1}_{-0.1}$ & 11.33$^{+0.001}_{-0.001}$ & 0.06$^{+0.01}_{-0.01}$ & 84.3$^{+0.3}_{-1}$ & -66$^{+0.24}_{-0.20}$ & 16  \\ \hline
\hline
\end{tabular}%
\end{adjustbox}
\end{center}
\caption{Best-fit parameters as obtained with \texttt{galario}, for each source, ALMA Band, and model. Sources with a {\large{(\textcolor{red}{$^{*}$})}} apex lack envelope continuum excess at either 3.1 mm or both wavelengths and they were modeled using one or more compact components. See section \ref{methods_faust} for more details. The cases in which the \protect$\chi^2$ value is significantly higher than 1 are the ones for which high signal-to-noise asymmetries highlight the limitation of a spherically symmetric model (e.g., IRAS4A, VLA1623A, L1551-IRS5) or where secondary sources in the FOV are left in the residuals (IRS7B, GSS30, IRAS4C; see text in Section \ref{methods_faust}).} 
\label{tab:fits_results}
\end{table*}

For the reasons above, we mainly conduct our analysis in the uv-plane, where we aim to model the data with multiple components that trace the inner compact regions as well as the large-scale envelopes. This approach has been benchmarked by e.g., \citet{Maury2018}, \citet{Galametz2019}, \citet{Cacciapuoti2023}, and \citet{Tung2024}. The main goal of the modeling is to isolate the central compact, usually optically thick, component. Once achieved, this compact component can be subtracted from the data, so that only the envelope flux contribution remains, which can then be used to compute an uncontaminated spectral index. It is noteworthy that this uv-plane approach is not free of shortcomings, the main one being the loss of information about the 2D source morphology: when computing $\alpha$ as a function of azimuthally averaged baseline bins, we effectively average the envelope properties in azimuth as well. However, this is a small price to pay when considering that (i) for the more compact objects it is not possible to generate an uncontaminated (from the disk) 2D $\alpha$ map, (ii) at this stage we are interested in the general trend of the spectral index across the envelopes rather than their precise 2D distribution.  

To fit the continuum emission visibilities of a system made of $n$ components, we consider the total model to be the sum of the visibilities of the individual components. The number of parameters in our models varies based on the number of components needed to fit the data as decided based on the source morphology: e.g., a power-law envelope and/or one or more Gaussian disks. In particular, lacking a general description for the often irregular morphologies of collapse, the brightness profile of typical protostellar envelopes has long been described by Plummer profiles (\citealt{Plummer1911}, \citealt{Motte2001}, \citealt{shirley2002}, \citealt{Maury2019}):
\begin{equation}
    \textrm{I$^{\textrm{P}}_{\textrm{TOT}}$(R)} =  \frac{\textrm{I$^P_0$}}{\Bigg[1 + \Big(\frac{\textrm{R}}{\textrm{R$_i$}}\Big)^2 \Bigg]^{\frac{\textrm{p+q}-1}{2}}} .
    \label{eq:total_profile}
\end{equation}
This envelope power law has peak flux \textrm{I$^{P}_0$} and three other free parameters: \textrm{R$_{\textrm{i}}$}, \textrm{R$_{\textrm{out}}$}, \textrm{p+q}. These are such that the envelope emission is constant within a radius \textrm{R$_{\textrm{i}}$} to avoid divergences, it follows a power law between \textrm{R$_{\textrm{i}}$} and \textrm{R$_{\textrm{out}}$}, and it is null beyond \textrm{R$_{\textrm{out}}$}. The power law index of the brightness profile \textrm{p+q} is a combination of the power laws indices that describe the envelope density (\textrm{R$^{\textrm{-p}}$}) and temperature (\textrm{R$^{\textrm{-q}}$}).

For the inner optically thick regions, dominated by a combination of disk and free-free emission, we instead use a 2D Gaussian model (and a point source where necessary). The Gaussian is defined by a peak \textrm{I$^G_0$}, a width $\sigma$, an inclination ``inc'' and position angle PA, so that it can describe elliptic disks:
\begin{equation}
\textrm{I$^{G}_{\textrm{TOT}}$(R)} = \textrm{I$^G_0$} \cdot e^{-\frac{\textrm{R}^2}{2\sigma^2}}
\end{equation}

In order to fit these models, we used \texttt{emcee} \citep{Foreman-Mackey2013} and \texttt{galario} \citep{Tazzari2018}. The former can be used to sample the parameter space using a Markov Chain Monte Carlo approach, in order to set a combination of parameters at each step. The latter computes a model image given a brightness profile and the parameters as selected by \texttt{emcee} at each iteration. Then it Fourier-transforms the model into synthetic visibilities and samples them at the uv-points covered by the antenna configurations with which the observations were performed. Finally, it runs a minimum-$\chi^2$ test between the data and the model visibilities at each trial of parameters sets, until it converges to a minimum.

\begin{table}[t]
\begin{center}
\resizebox{\linewidth}{!}{\begin{tabular}{c|c|c|c|c}
    \hline \hline
        Source & F$_{\textrm{3.1 mm}}$ (mJy) & F$_{\textrm{1.2 mm}}$ (mJy) & FWHM (au) & $\alpha^{\textrm{compact}}$ \\
        \hline
        IRAS15398-3359 & 1.28 & 8.09 & 9 & 2.2 $\pm$ 0.2 \\
        VLA1623A & 21 & 150 &  111 &2.3 $\pm$ 0.2 \\
        Elias 29 & 3.73 & 14.1 & 9 & 1.6 $\pm$ 0.2 \\
        CB68 & 7.1 & 52 & 26 & 2.1 $\pm$ 0.3 \\
        BHB07-11 & 10.99 & 114.5 & 93 & 2.7 $\pm$ 0.2 \\
        L483 & 2.32 & 17.4 & 23 & 2.4 $\pm$ 0.2\\
        IRAS4A1 & 25 & 280 & 88 & 2.6 $\pm$ 0.2 \\
        GSS30 & 21.8 & 106.8 & 90 & 1.9 $\pm$ 0.2 \\
        IRS7B & 48.2 & 282.1 & 55 & 2.0 $\pm$ 0.2\\
        IRS63 & 37.2 & 304.1 & 72 & 2.3 $\pm$ 0.2 \\
        IRAS4C & 9.6 &  67.4 & 95 & 2.2 $\pm$ 0.2 \\
        L1551-IRS5 & 43 & 300 & 31 & 2.1 $\pm$ 0.2  \\
        L1527$^{(a)}$ & 28.0 & 132.0 & 75 & 2.1 $\pm$ 0.2 \\
        \hline
    \end{tabular}}
    \end{center}
    
    \caption{The 3.1 mm and 1 mm fluxes of the fitted compact component for each source. Almost all spectral indices (third column) are consistent with optically thick emission (see text). Note: (a) is from \citet{Cacciapuoti2023}.}
    \label{tab:compact_sources_alpha}
\end{table}

The parameters were all set to vary within uniform prior distributions. The offsets from phase centre of each fitted component, $\Delta$RA and $\Delta$Dec, were set tightly (a few tenths of arcseconds ranges) around values obtained with an inspection of data with the Cube Analysis and Rendering Tool for Astronomy (CARTA, \citealt{Comrie2021}); the Gaussian width $\sigma$ were set to vary between zero and 0\farcs7 (roughly a 2$\sigma$ 200 au disk at 140 pc), which is wide enough to accommodate any known Class 0/I disk (e.g., \citealt{Maury2022}); the inclination and PA were set to vary over the full possible ranges. The Plummer envelope's inner radius varied between zero and 2\farcs0 (in order to ensure it to be at least as extended as the disk); the outer radius \textrm{R$_{\textrm{out}}$} would vary between 2\farcs0 and 15\farcs0 (slightly larger than the maximum recoverable scale of the observations). Finally, the envelope power law exponent \textrm{p+q} was set free in the range [1,3]. We ran 120 walkers for 25000 steps to reach convergence and samples the posterior distribution of each fit.
\begin{figure*}[t]
    \centering
    \includegraphics[width=\linewidth] {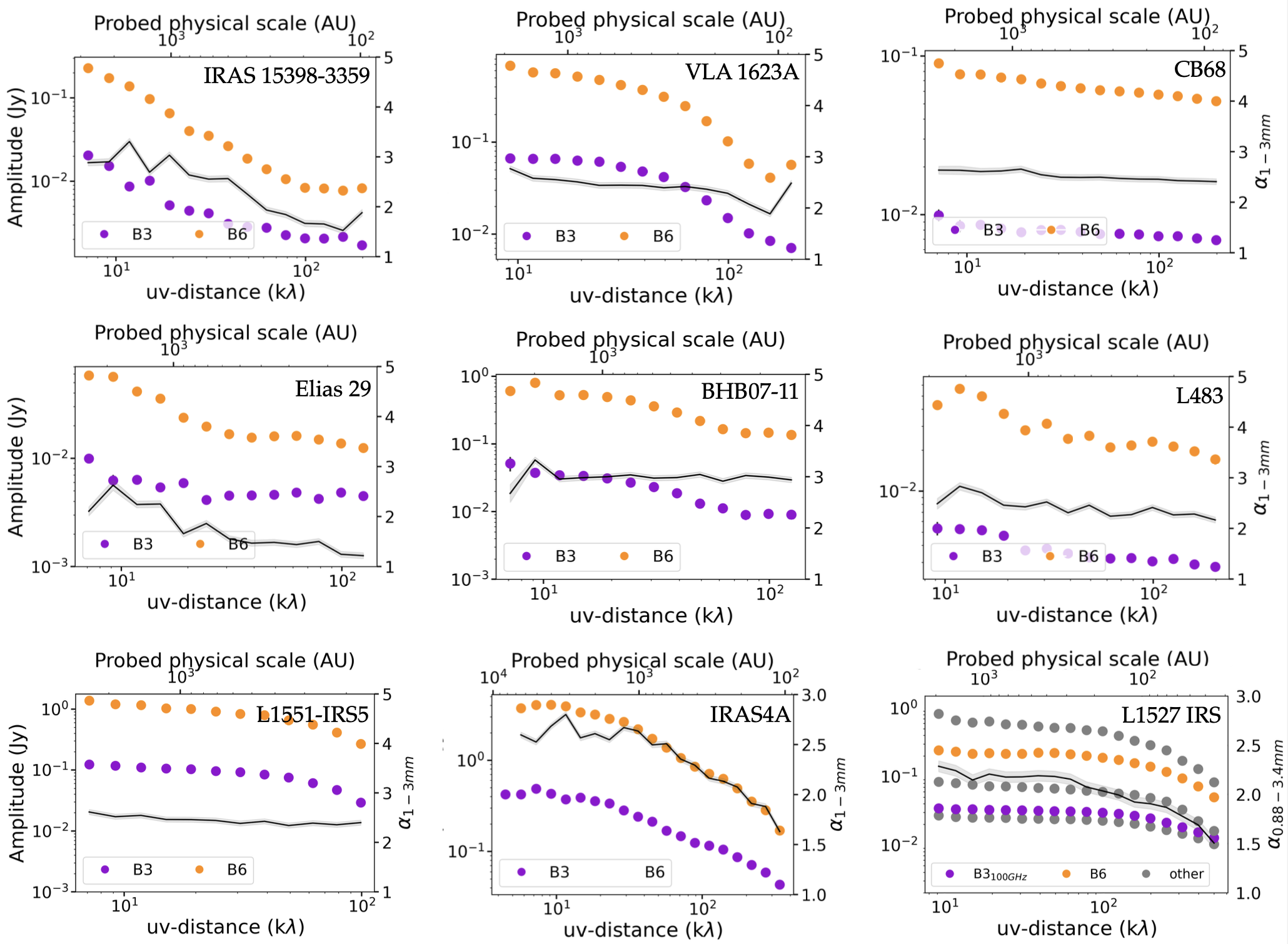}
    \caption{The binned azimuthally averaged amplitude profiles in orange (B6) and violet (B3) for each source refer to the left y-axis of each panel.  
    The total emission spectral index (black line) of the emission as a function of uv-distance (bottom) or, equivalently, probed physical scale $\theta = 1.22 \lambda /$uv-distance (top), refers to the right y-axis of each panel.
    Note: L1527 IRS is from \citet{Cacciapuoti2023}, where only 1.2 and 3.1 mm wavelengths (similar to FAUST) have been colored accordingly to the other sources, while the other frequencies are in gray.}
    \label{fig:alphas}
\end{figure*}

We note that when more than one source was present in the field of view, we modeled and subtracted those far away from the target from the visibilities in CASA (using the \texttt{uvmodelfit} and \texttt{uvsub} routines). Only after this step we modeled the residual visibilities with \texttt{galario}. If, instead, any secondary source was close to the target source (blended with envelope emission of the primary target), we modeled the two self-consistently in \texttt{galario} in order to avoid subtracting flux from the close envelope with a rough CASA model.
The sources in our sample with secondary emission in the field of view are GSS30, RCra IRS7B, VLA1623A and IRAS4A. GSS30 and RCra IRS7B do not show any sign of extended envelopes (in one or both bands) and thus fall out of the sources of interest for this study: for these we do not model out the secondaries since we do not measure an envelope $\beta$ in any case. The spectral index of the fitted compact component for those sources is thus contaminated by these secondaries. However, the flux of the primary targets always far exceeds the secondaries (by $\gtrsim 10$ times) and thus would not change our conclusions: these protostars host compact, optically thick disks with $\alpha \sim 2$. More generally, the sources consistent with no envelope emission in at least one band - and so for which no $\beta$ was measured - are: GSS30, RCra IRS7B, IRS63.
\begin{figure*}[t]
    \centering
    \includegraphics[width=\linewidth] {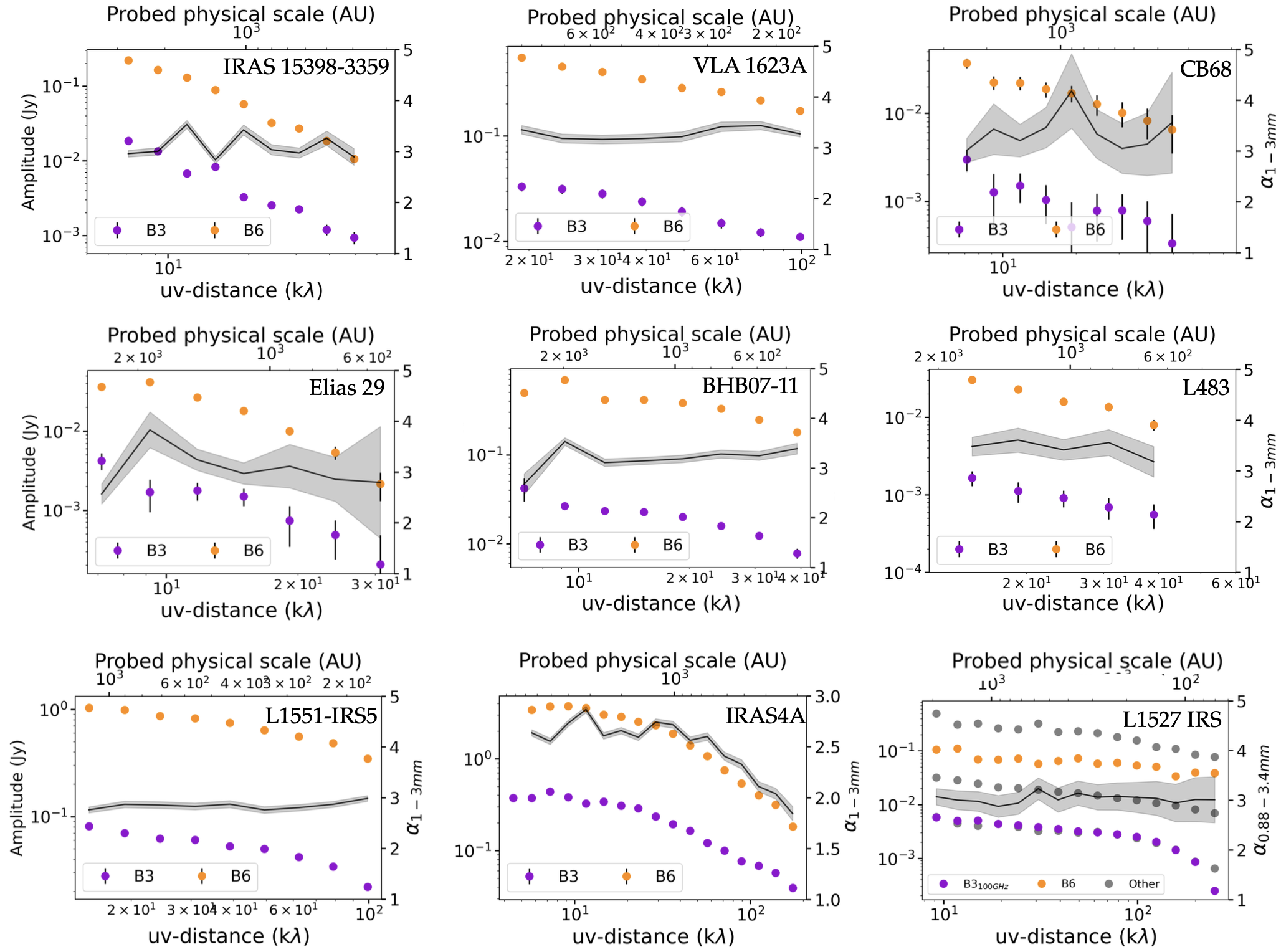}
    \caption{The binned azimuthally averaged amplitude profiles after subtraction of the model compact component, hence of the envelope flux alone, in orange (B6) and violet (B3) for each source refer to the left y-axis of each panel.  
    The envelope spectral index (black line) as a function of uv-distance (bottom) or, equivalently, probed physical scale $\theta = 1.22 \lambda /$uv-distance (top), refers to the right y-axis of each panel.
    Note: L1527 IRS is from the multi-wavelength analysis of \citet{Cacciapuoti2023}, where only FAUST-like wavelengths have been colored uniformly to the other sources and the other frequencies are in gray.}
    \label{fig:alphas_env}
\end{figure*}
On the contrary, the secondary field of view sources around VLA1623A and IRAS4A need to be treated more carefully since we could derive envelope $\beta$ measurements for them. 
For the former, after subtraction of off-centre sources in the field of view (Fig. \ref{fig:vla1623a_remove}), the data was modeled as a power law for the envelope and two Gaussians for the central binary (Fig. \ref{fig:vla1623_fit}). 
For the latter, after removal of secondaries (Fig. \ref{fig:iras4a_remove}), the best fit consisted of one Gaussian component for IRAS4A1 and two Plummer envelopes, as extended emission is clearly detected with high signal to noise ratio around both protostars (Fig. \ref{fig:sample_1.2 mm}). Previously, \citet{Maury2018} found that only two Plummer profiles would suffice to well fit the PdBI $+$ IRAM 30m telescope CALYPSO data and derived similar values for the envelope profile parameters. The higher resolution of our data cause the visibilities profile to drop faster at the long baselines with respect to what observed in that previous study, and this drop is better fit by an additional component with $\sim$90 au in size. As no disk kinematic signature has been detected down to 50 au for this source, we interpret this compact emission to be due to the very inner, optically thick envelope.
\begin{figure*}[t]
    \centering
    \includegraphics[width=\linewidth] {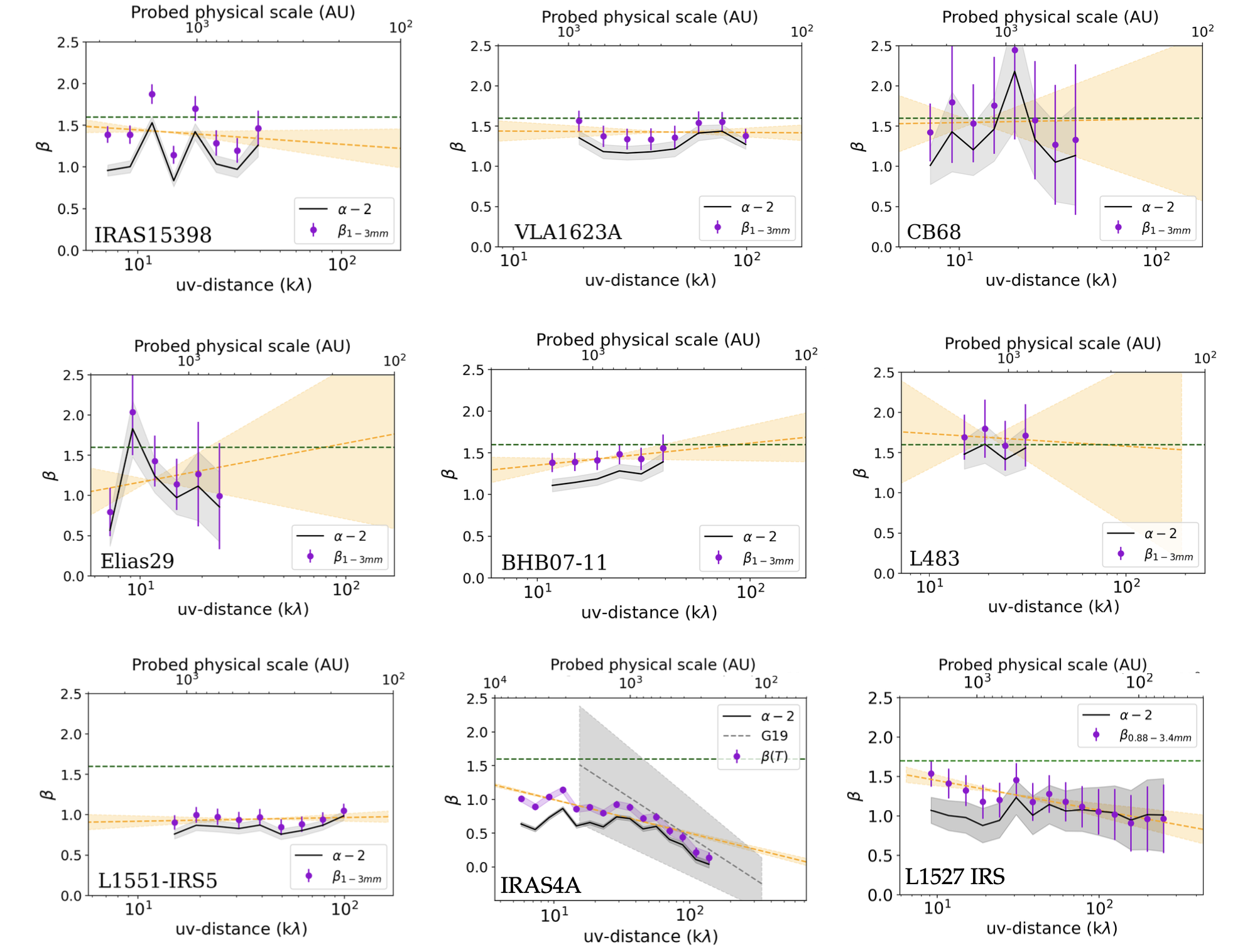}
    \caption{The derived dust opacity spectral indices for the sources studied in this work, as a function of uv-distance (bottom) or traced physical scales (top). The black lines represent the $\beta = \alpha - 2$ approximation, while the purple dots are derived as $\beta = \alpha - \textrm{d}\log \textrm{B(T)}/ \textrm{d}\nu$ in each bin. The orange line is a linear fit to the purple points. The green dashed line is the typical ISM $\beta \sim 1.6$ value. 
    Notes: (i) some sources display more bins than others due to differences in the signal-to-noise ratio of the detection; (ii) for IRAS4A we also report the measured $\beta$ profile of \citet{Galametz2019} (gray dashed line and 1$\sigma$ shaded region), limited within the scales that their data probed; and (iii) the profile for L1527 IRS is from \citet{Cacciapuoti2023}.}
    \label{fig:betas}
\end{figure*}
Finally, we note that two Gaussians were found to fit BHB07-11 (Fig. \ref{fig:bhb_fit}) better than a Gaussian plus Plummer model. This is not surprising when inspecting the image plane, in which this source shows a flattened, disk-like inner envelope that extends for approximately 500 au (Fig. \ref{fig:sample_1.2 mm}). 

The best fit parameters are summarised in Table \ref{tab:fits_results} and the obtained profiles are overplotted on the data in appendix \ref{app:fits}. The fit $\chi^2$ is also reported: in some cases, the $\chi^2$ results significantly higher than unity. This captures the cases for which strong asymmetries characterize the envelopes and are not well described by a spherical model (e.g., IRAS4A, VLA1623A), or the cases for which unsubtracted FOV sources remain in the residuals (IRS7B, GSS30, IRAS4C). In both cases, we reaffirm that the main objective of the fits is to describe the inner optically thick region to subtract it from the data and thus measure the envelope uncontaminated spectral index, and not to perfectly model the entirety of the very irregular emission of protostellar envelopes.

It is noteworthy that, in almost all cases, the Gaussian emission contribution reaches its maximum in the uv-space (integrated flux) at about 150-200 k$\lambda$, the same baselines at which \citet{Galametz2019} considered the compact emission to saturate (based on the disk sizes found for the same sample in \citet{Maury2018}) and hence the one they subtracted at shorter baselines to isolate the envelope contribution from the compact central regions in the uv-space. We thus confirm that their approach is robust in what regards the subtraction of the compact component from the visibilities before measuring $\alpha$ and $\beta$.

\section{Spectral index profiles}
\label{results_alpha_beta}

The spectral indices were computed as:  
\begin{equation}
    \alpha = \frac{\log F_{\nu_2} - \log F_{\nu_1}}{\log \nu_2 - \log \nu_1},
    \label{eq:alpha}
\end{equation}
where F$_{\nu_i}$ are the fluxes at the two observed ALMA bands, represented by their mean frequencies $\nu_1 = 95$ GHz and $\nu_2 = 233$ GHz. The quantity in Eq. \ref{eq:alpha}, was computed in two ways for each source. In addition to the statistical uncertainty, all computed spectral indices in this paper carry a systematic 1$\sigma$ error of 0.1 from flux calibration. This flux calibration error corresponds to 2.5\% and 5\% for Band 3 and Band, 6 respectively (see ALMA Cycle 6 Technical Handbook).

First, a profile of the spectral index as a function of the uv-distance was computed starting from the total emission (Fig. \ref{fig:alphas}). Since both disk and envelope contribute at all uv-distances, this spectral index is not representative of one or the other individually, however, it mainly traces the disk emission for most sources in which the compact emission is dominating in flux, especially at long baselines. Subsequently, we show (Fig. \ref{fig:alphas_env}) the $\alpha$ profile after the subtraction of the compact region emission (modeled as explained in Section \ref{methods_faust}): this represents the spectral index of the envelope emission, in which we are mainly interested.

In Fig. \ref{fig:alphas}, the spectral index profiles of the total emission are mostly consistent with the spectral index of the fitted compact component, as presented in Table \ref{tab:compact_sources_alpha}. Almost all spectral indices obtained for the total emission result around 2, except the ones of BHB07-11 ($\alpha \sim 2.7$, consistent with \citealt{Alves2018}) and Elias 29, for which a low $\alpha \sim 1.6$ is derived and could be caused by e.g., scattering effects \citep{Liu2019}. This means that the high signal-to-noise disks are dominating the total flux and are described by low spectral indices, likely due to their high optical depth.

In Fig. \ref{fig:alphas_env}, where the spectral index characterizes the emission from the envelope alone, one can appreciate how $\alpha$ is significantly larger than the total emission case for any given source. This happens because the contamination of the optically thick component has been disentangled.

Finally, we computed the dust opacity spectral index $\beta$ in the optically thin and Rayleigh-Jeans approximation, $\beta = \alpha - 2$, and in its non-approximated form $\beta = \textrm{d}\log \tau / \textrm{d}\nu = \alpha - \textrm{d}\log \textrm{B(T)}/ \textrm{d}\nu$, where we had to assume a temperature profile. Based on, e.g. \citet{Adams1985}, \citet{Motte2001}, we parametrised the dust temperature as a function of distance from the protostar and its luminosity as follows: 
\begin{equation}
    \textrm{T(R)} = 38 \times \Bigg(\frac{\textrm{L}}{\textrm{L$_{\odot}$}}\Bigg)^{0.2} \Bigg(\frac{\textrm{R}}{100 \textrm{au}}\Bigg)^{-0.4} \textrm{K}.
    \label{eq:temp}
\end{equation}
The resulting optical depths are always much lower than unity, except for the most massive envelope of IRAS4A for which it reaches levels of the order of few tenths. The temperature power law we consider (cfr. Eq. \ref{eq:temp}) yields temperatures of about 25 K at 200 au. At these temperatures, $\beta \sim \alpha  - 1.85$. Even by varying the reference temperature (T $=$ 38 K at 100 au) in Eq. \ref{eq:temp} by 10 K, or by varying the power law index of the temperature profile in the range [-0.3,-0.6], the derived $\beta$ values remain well within a 10\% uncertainty.
The resulting profiles of $\beta$ are shown in Fig. \ref{fig:betas}, limited in a range of baselines over which the envelope contribution to the total emission is significant\footnote{After this range, towards longer baselines, the disk emission largely dominates and its subtraction translates into large error bars on the remaining flux bins.}. The uv-plane gradient of the profile in the envelope is simply defined by:
\begin{equation}
    \beta^{\textrm{env}}(\textrm{R}_{\textrm{au}}) = \beta^{\textrm{env}}_{\textrm{grad}} \cdot \log_{10}( \textrm{R}_{\textrm{au}}) + \textrm{a}.
    \label{eq:beta_gr}
\end{equation}
We note that for IRAS4C we placed lower limits to the spectral index $\alpha$ and the dust emissivity spectral index $\beta$, since the emission was only detected at 1.2 mm (see also section \ref{sec:cavities}). Moreover, for RCrA-IRS7B we could only derive a spectral index in the image plane using the maps of Fig. \ref{fig:sample_1.2 mm} and Fig. \ref{fig:sample_3.1 mm}: only using all the available baselines could we detect the faint, extended emission at 3.1 mm, lost in a baseline-binned approach. However, since the extended emission is quite far from the centre, the contamination from it should be negligible (Fig. \ref{app:2dmaps}).

\subsection{No envelope excess for the Class I sources}
\label{sec:no_excess}

Among the studied sources, two present no envelope excess at short spacings and are well fit by a compact component alone. These are: GSS30 (Fig. \ref{fig:gss30_fit}) and IRS63 (Fig. \ref{fig:irs63_fit}). These sources are not discussed further since we are here interested in envelopes properties. The main reason for the lack of a clear detection likely stands on the ground of the relatively poor sensitivity of the data in combination with the evolutionary stage of these sources, for which the envelope does not represent a dominant component of the system anymore. 
When comparing to available literature values for the retrieved disk radii of these sources, we find consistent results. Our fits yield a FWHM radius of 90 au for the disk of GSS30, which is consistent with the reported one from high resolution 1.2 mm observations of the ALMA eDisk Large Program (98 au; \citealt{Santamaria-Miranda2024}). The fitted disk size of IRS63 is 78 au, in agreement with the continuum observations of \citet{Segura-Cox2020}. It is worth mentioning that, using the integrated Gaussian component fluxes output by our modeling scheme, we derive spectral indices consistent with optically thick emission for both disks ($\alpha \sim 2$, Table \ref{tab:compact_sources_alpha}). These values do not exclude the possibility of grain growth in these young disks, but more detailed conclusions about it are out of the scope of this work.


\subsection{The dust emissivity index across FAUST envelopes}
\label{sec:beta_class0}

For 11 FAUST sources we detect envelope emission in both bands (except for IRAS4C, only detected at 1.2 mm) and thus we are able to perform our analysis and measure dust opacity spectral indices (one is from the pilot study of \citealt{Cacciapuoti2023}).
We find that for BHB07-11, VLA1623A, CB68, L483, IRAS15398-3359, Elias 29, the 500 au dust opacity spectral index profile is consistent with a flat line within 1$\sigma$ error bars, and with values $\beta \sim 1.4 - 1.7$. While a full profile could not be obtained for IRAS4C, a significant lower limit, $\beta > 1.4$, was placed. These $\beta$ values are thus all consistent within error bars with typical ISM values, indicating maximum grain sizes of the dust distribution well below the suggested grown grains of $\sim$100 $\mu$m of previous studies (\citealt{Miotello2014}, \citealt{Galametz2019}, \citealt{Cacciapuoti2023}). 

On the other hand, three sources have $\beta \lesssim 1$ in at least some range of physical scales. First, the dust emissivity spectral index of L1527 IRS varies in the range [1, 1.5] is discussed in the pilot study by \citet{Cacciapuoti2023}. Second, IRAS4A displays a gradient in $\beta$ in the range [0, 1] (see section \ref{sec:iras4a}). Finally, L1551-IRS5 is characterized by a flat $\beta \sim 1$ profile (see section \ref{sec:cavities}. The $\beta$ values for these three sources are thus consistent with larger-than-ISM grains.

In Table \ref{tab:all_betas}, we report the measured $\beta$ gradients and their values at a representative scale of 500 au, to enable comparison to the previous sample study of \citet{Galametz2019}.

\begin{figure}[t]
    \centering
    \includegraphics[width=\linewidth] {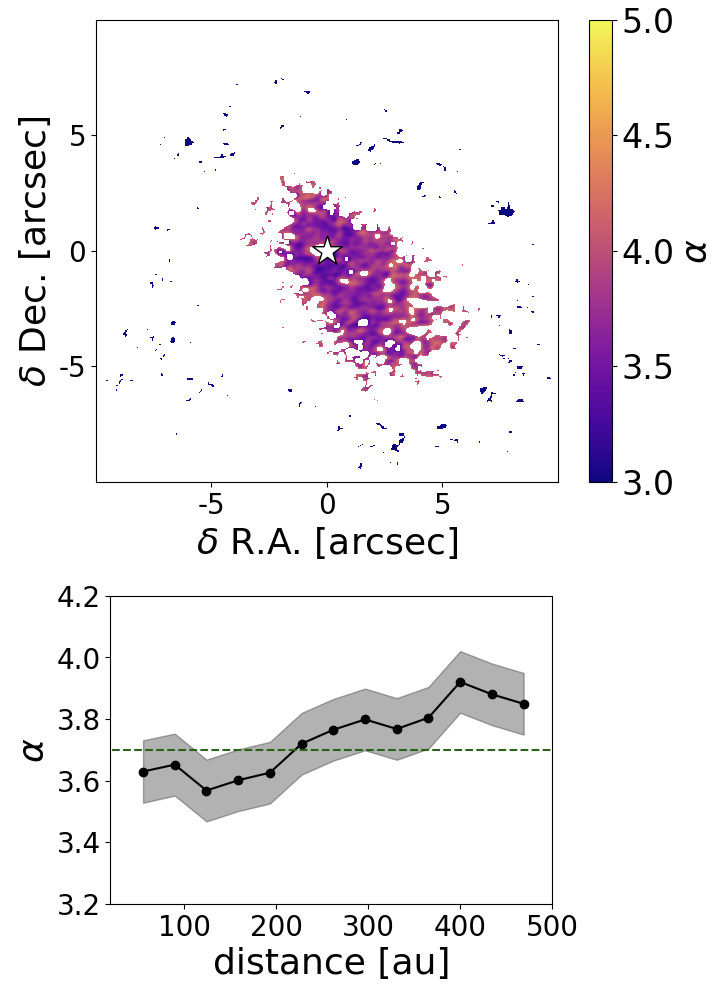}
    \caption{A 2D map of the spectral index $\alpha$ around the protostar IRAS15398-3359, the position of which is shown by a white star. Pixels are non-zero where both 1.2 mm and 3.1 mm emission is present at 3$\sigma$ level, at least. The lower panel profile has been obtained by azimuthally averaging over the map, between 50 and 500 au (disk radius is $\sim$9 au). It shows $\alpha \sim 3.7$ along the dusty cavity walls of IRAS15398-3359. The dashed green line represents ISM-like values.}
    \label{fig:iras15398_alpha_map}
\end{figure}

As anticipated in Section \ref{methods_faust}, computing 2D spectral index maps is a challenging task and their interpretation is limited by the mutual contamination of disk and envelope. However, in cases of high signal-to-noise ratio of the extended emission, an attempt can be made since the disk emission will only contaminate up to certain scales, after which the envelope spectral index becomes almost pure. In order to better resolve disk from envelope and compute 2D $\alpha$ maps, we re-imaged the data for this sources to obtain beam-matching maps and using the full extent of the visibilities (as opposed to what explained in section \ref{obs_faust} and displayed in Fig. \ref{fig:sample_1.2 mm} and Fig. \ref{fig:sample_3.1 mm}). The maps statistics for this analysis are in Table \ref{tab:2d_alpha_tab}. The sources for which we were able to produce meaningful spectral index maps are: BHB07-11, VLA1623A, IRAS4A, IRAS15398-3359, and L1551-IRS (Fig. \ref{fig:iras15398_alpha_map} and Fig. \ref{fig:2d_maps}). All show spectral indices which are consistent with the ones of the compact emission reported in Tab. \ref{tab:compact_sources_alpha} within $\lesssim$200 au, and consistent with the envelope ones reported in Tab. \ref{fig:alphas_env} at $\sim$500 au where the contamination of the inner emission becomes negligible. Finally, these image plane spectral indices are consistent with the ones computed in the uv-plane analysis.

\begin{table}[t]
    \centering
    \begin{tabular}{c|c|c}
    \hline \hline
        Source & $\beta^{env}_{grad}$ & $\beta^{env}$ (500 au)  \\
        \hline
        IRAS15398-3359 & -0.17 $\pm$ 0.20  &  1.41 $\pm$ 0.09 \\
        VLA1623A & 0.45  $\pm$ 0.25 & 1.43  $\pm$ 0.05  \\
        CB68 & 0.07  $\pm$  0.84 & 1.58 $\pm$ 0.42  \\
        Elias 29 & 0.48 $\pm$ 0.99 & 1.42 $\pm$ 0.50  \\
        BHB07-11 & -0.26 $\pm$  0.30 & 1.50 $\pm$ 0.08   \\
        L483 & -0.15 $\pm$ 0.79 &  1.69  $\pm$ 0.40 \\
        IRAS4A & 0.76 $\pm$ 0.11 & 1.10 $\pm$  0.03 \\
        IRAS4C &   -  &  $>$ 1.4    \\
        L1551-IRS5 & -0.04 $\pm$ 0.11 & 0.94 $\pm$ 0.04 \\
        RCrA-IRS7B & - & 3.3 $\pm$ 0.1 \\
        L1527$^{(a)}$ & 0.39 $\pm$ 0.17 & 1.24 $\pm$ 0.06 \\
        \hline
    \end{tabular}
    \caption{For each source for which extended emission was detected in at least a band, we report the measured gradient of the dust opacity spectral index $\beta^{env}_{grad}$ (Eq. \ref{eq:beta_gr}) and a representative value of $\beta^{env}$ at 500 au, after subtraction of the central compact component. The value of $\beta$ is extrapolated from the uv-plane for all sources except for IRAS4A, where the image plane value was used (see section \ref{sec:iras4a}). Since IRAS4C is detected only at 1.2 mm, a lower limit is derived. For RCrA-IRS7B, the spectral index is derived from the image plane (see section \ref{results_alpha_beta}). Note: (a) is from \citet{Cacciapuoti2023}.}
    \label{tab:all_betas}
\end{table}

\subsection{The massive envelope of the IRAS4A proto-binary}
\label{sec:iras4a}

We fitted the IRAS4A source with a model that accounts for two disk-like components and two Plummer profiles, in order to capture the binary nature of this system (Fig. \ref{fig:sample_1.2 mm}). 
The resulting $\beta$ profile in Fig. \ref{fig:betas} is obtained after subtraction of the two fitted compact components, and thus traces a combination of the remaining contribution of the envelopes. The measured dust opacity spectral index is consistent with the one found by \citet{Galametz2019}, varying between 0.5 and 1.1 across the same range of scales. However, by extending the available probed scales in this work, we can see how the gradient in the uv-plane profile ($0.76 \pm 0.11$) is shallower than the one reported by them ($1.31 \pm 0.36$), because we are able to observe how $\beta$ flattens around 1.1 at shorter and shorter uv-distances (see IRAS4A panel in Fig. \ref{fig:betas}).

Since the extended envelope of IRAS4A appears very bright even at the high resolutions probed by the FAUST observations, we could also produce a resolved map of its spectral index (Fig. \ref{app:2dmaps}). 
One can observe how the spectral index varies from $\alpha \sim 2.6$ in the inner 200 au all the way to $\alpha \sim 3.1$ at 500 au from IRAS4A1. These values are consistent with the ones we find the in the uv-plane, but seemingly at different scales. This is the case simply because the correspondence between the uv-distance and the on-sky distances is not one-to-one. In the uv-plane, shorter and shorter uv-distances recover the \textit{integrated} flux of larger and larger sky portions. This means that the short baselines fluxes and spectral indices are integrated results of what is contained within a certain scale. After contribution of the central component, a bright envelope can still play this role. And while this is true for any source, it becomes especially important for IRAS4A because of its outstanding brightness.


The $\alpha$ map of IRAS4A shows how the innermost regions of the envelopes, centered on IRAS4A1 and IRAS4A2, are characterized by a spectral index $\alpha \sim 2.6$. Further and further away from the centre of the sources, the spectral index increases and reaches $\alpha \sim 3.1$ at 500 au away from IRAS4A1, consistently with the profile measured in the uv-plane (Fig. \ref{fig:alphas_env} and Fig. \ref{fig:betas}). 
Finally, as the temperatures obtained with Eq. \ref{eq:temp} are relatively high for IRAS4A even at 500 au (T$>$50 K), $\beta$ can be approximated as $\alpha - 2$. Thus, the envelope dust opacity spectral index $\beta \sim 0.6-1.1$ between 50 and 500 au. 


The reader interested in the very inner 100 au of IRAS4A can refer to \citet{Guerra-Alvarado2024}. 

\begin{figure*}[h!]
    \centering
    \includegraphics[width=\textwidth] {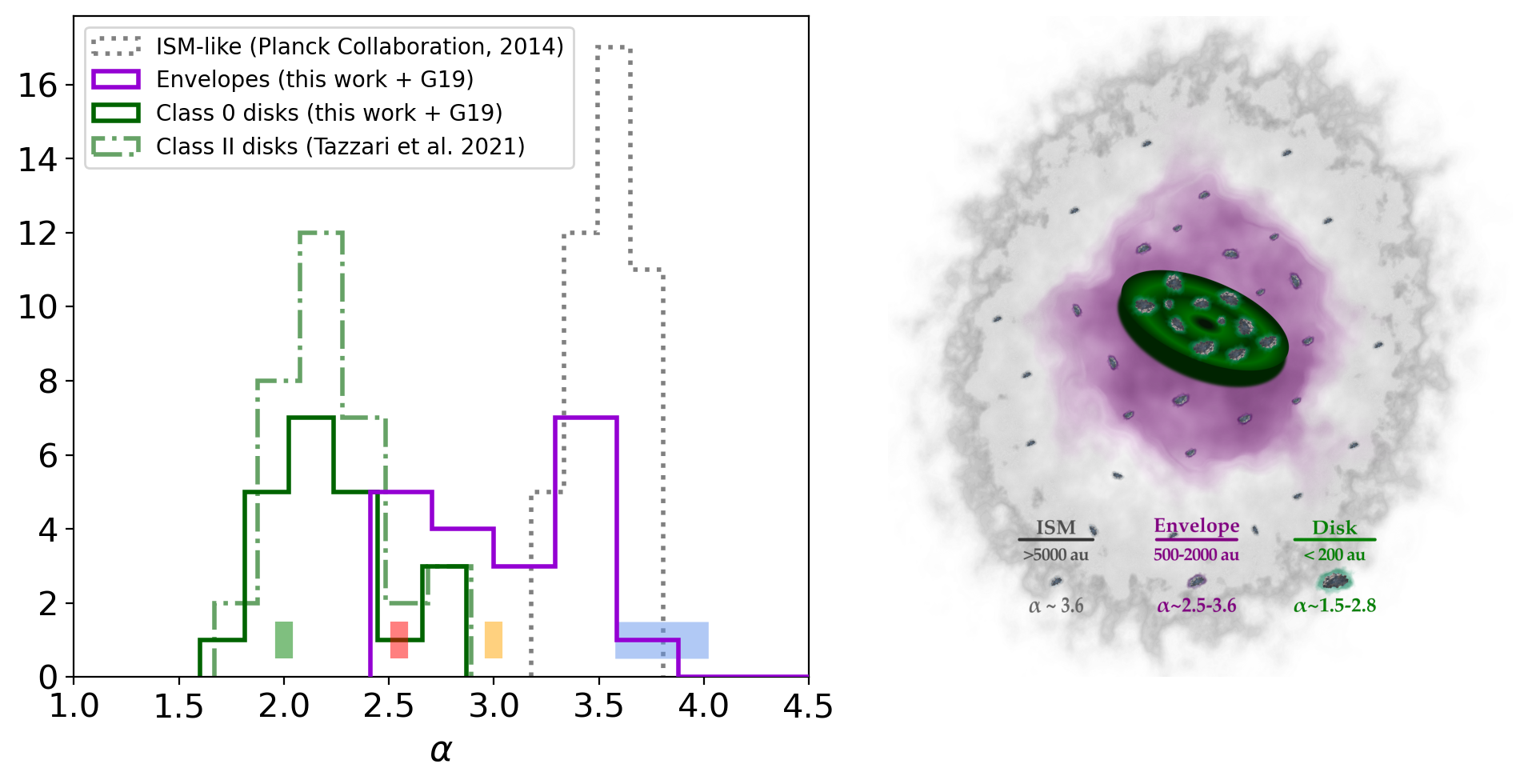}
    \caption{Spectral indices distributions ordered following spatial and time evolution of dust in star- and planet-forming environments: (i) $\alpha$ in the submillimetre ($<$350 GHz) for the diffuse ISM (grey dotted representative distribution from \citet{Planck2014}); (ii) at 1-3.1 mm for the protostellar envelopes at 500 au scales from this work and G19 combined (violet); (iii) at 1-3.1 mm for the class 0 protoplanetary disks from this work and G19 combined (green); (iv) for typical class II disks at 1-3.1 mm (green dash-dotted line, \citealt{Tazzari2021}). We also show the spectral indices reported by \citet{Bracco2017} for prestellar cores (blue box), Class 0 YSOs (red/orange boxes), and a T Tauri disk (green box), all within a common filament in Taurus.
    The spectral indices of protostellar envelopes appear to be bridging the gap between ISM and disks, hints to a continuous evolution of dust properties from one to the other. On the right, a sketch represents the evolution of dust that emerges from the spectral indices variations, with growth from the (sub-)micron grains in the ISM to the mm/cm-sized grains of disks, through infalling envelopes. The typical scales and values are the ones of the aforementioned studies.}
    \label{fig:alpha_disks_envs}
\end{figure*}
\subsection{Dust emission along outflow cavity walls: IRAS15398-3359, IRAS4C, Elias 29, RCrA-IRS7B, L1551-IRS}
\label{sec:cavities}
We here discuss peculiar cases in terms of morphology, where dust emission is recovered along cavity walls carved by protostellar outflows.

The protostar IRAS15398-3359 displays dust emission along the cavity walls that its outflows are carving in the collapsing envelope (Fig. \ref{fig:sample_1.2 mm}), as already observed by \citet{Jorgensen2013} at 0.88 mm. With the ALMA FAUST data, we can now measure the dust opacity spectral index as the cavity walls are clearly detected at both 1.2 and 3.1 mm, which results to be $\beta \sim 1.45 \pm 0.10$ on average, consistent with small dust grains. Thanks to the relatively high signal-to-noise ratio and the very extended nature of the emission around this source, we could produce a resolved map of the spectral index which is expected to be uncontaminated from the disk far away from the centre (since the disk is very compact and the rest of the emission is very extended). We find that it is well consistent with typical ISM values, $\alpha \sim 3.6$. If one considers that $\beta = \alpha -2$ as a first approximation, this values of $\alpha$ result consistent with our uv-plane analysis and with dust grains sizes well below 100 $\mu$m (Fig. \ref{fig:iras15398_alpha_map}). In this case, the Rayleigh-Jeans approximation $\beta = \alpha -2$ works well since the cavity walls of outflows are at higher temperatures than cold envelopes (\citealt{Visser2012}, \citealt{Lee2015}).

The morphology of the dust continuum of IRAS4C follows closely the morphology of the outflow cavity walls seen in, e.g., CS and CCH \citep{Zhang2018}. As the extended emission is only detected at 1.2 mm, we can only place a lower limit. Based on the 3$\sigma$ detection in Fig.\ref{fig:sample_1.2 mm} and on the rms of the 3.1 mm data (Tab. \ref{tab:map_stats}), find $\alpha > 3.4$. Thus, as for IRAS15398-3359, the spectral index of IRAS4C is consistent with ISM-like values and thus small dust grains.

The nature of the extended emission around Elias 29 also appears to be tightly linked to the S-shaped H$_2$ outflow detected by \citet{Bussmann2007}, rather than to an infalling envelope. As this outflow appears to be encountering an expanding bubble south of the source (\citealt{Johnstone2000}, \citealt{Boogert2002}), shocks might be brightening up the dust emission or accumulating large amounts of dust in a localised space \citep{Oya2025}. It is thus not trivial to exclude that the observed emission and its spectral index $\alpha \sim 3$ is a combination of extended dust thermal emission and free-free emission. We thus caution about the grain growth interpretation of \citet{Miotello2014}, who estimated a dust opacity spectral index $\beta = 0.6 \pm 0.3$ with lower resolution observations, consistent within 1$\sigma$ with the one found in this work in the same range of uv-distances after the subtraction of the compact region: $\beta = 1.0 \pm 0.3$ (Fig. \ref{fig:betas}). A more detailed interpretation of the continuum emission of this source lays out of the scope of this work.

Beyond the sources of which above, a lower limit for $\alpha$ has already been published by \citet{Sabatini2024} for the dusty cavity wall associated with the RCrA-IRS7B system, as they did not recover the 3 mm emission in their high resolution maps. In this work, by imaging the data with a restricted uv-range (section \ref{obs_faust}), we recover the larger scale emission associated with the outflow cavity walls of this protostar and measure a spectral index along them $\alpha = 3.3 \pm 0.1$. Once again, assuming relatively high temperatures along the cavity walls, $\beta = \alpha - 2= 1.3$. Such a value is neither clearly ISM-like nor definitely consistent with dust growth: further data is needed to make a more accurate measurement in this case.

Finally, a detailed analysis of the cavity wall of L1551-IRS5 is consistent with the presence of relatively large dust grains ($\beta \lesssim 1$) along its cavity walls \citep{Sabatini2025}.

We thus note that the ALMA FAUST program is proving crucial for determining the dust properties along the walls of protostellar outflow cavities. 

\section{Discussion}
\label{discuss_faust}
In Section \ref{sec:intro}, we discussed the important roles dust plays in star and planet formation and in the ways we observe this processes through the Galaxy. Those aspects justify the need to further constrain the dust size distribution in protostellar envelopes, and thus the following discussion.

\subsection{A variety of spectral indices in protostellar envelopes}

Large surveys have been conducted in planet-forming disks to measure their spectral indices and to understand when and where grain growth starts (e.g., \citealt{Ricci2010}, \citealt{Tazzari2021}). As planet formation has been shown to possibly onset at the earliest stages of the star formation process (\citealt{Manara2018}, \citealt{Sheehan2018}, \citealt{Tychoniec2020}, \citealt{Seguracox2020}), much more attention has been drawn upon the study of dust emission during the Class 0/I stages of protostellar evolution (e.g., \citealt{Ohashi2023}, \citealt{Maureira2024}). Not only has there been a shift in time to try and look for the first hints of dust growth, a shift to larger scales has also been suggested towards possible dust agglomeration in cores and envelopes (see section \ref{sec:intro} for more details). 

It is thus becoming crucial to constrain spectral indices distributions for protostellar envelopes, bridges through which ISM dust is funneled onto young circumstellar disks. In Fig. \ref{fig:alpha_disks_envs} we show the measured 1-3.1 mm spectral indices for evolved Class II disks in Lupus \citep{Tazzari2021}; the ones of young Class 0/I disks from this work and \citet{Galametz2019}; a representative distribution ($\beta = 1.51 \pm 0.13$) of the ones at galactic scales as measured by \citet{Planck2014} as computed based on sub-mm wavelengths corresponding to 100, 143, 217, and 353 GHz frequencies; the spectral indices reported by \citet{Bracco2017} for prestellar cores, Class 0 YSOs, and a T Tauri disk within a common filament; and finally the 1 - 3.1 mm spectral indices of protostellar envelopes known to date (this work and \citealt{Galametz2019}). 

The distribution of the spectral indices displayed in Fig. \ref{fig:alpha_disks_envs} in the envelopes of the Class 0/I sources, in Class II sources and in the ISM distribution suggests a continuous evolution from more diffuse to denser media, explored for the first time at these intermediate scales.
This finding reinforces the idea of an evolutionary bridge between these types of environments, it extends to smaller scales the similar conclusions previously drawn at larger scales of \citet{Bracco2017}, i.e., between 4000 and 20,000 au, see Fig. \ref{fig:alpha_disks_envs}, and provides an important foundation for further investigation of dust evolution in its pathways from diffuse to denser media. 

\begin{figure}[t]
    \centering
    \includegraphics[width=\linewidth] {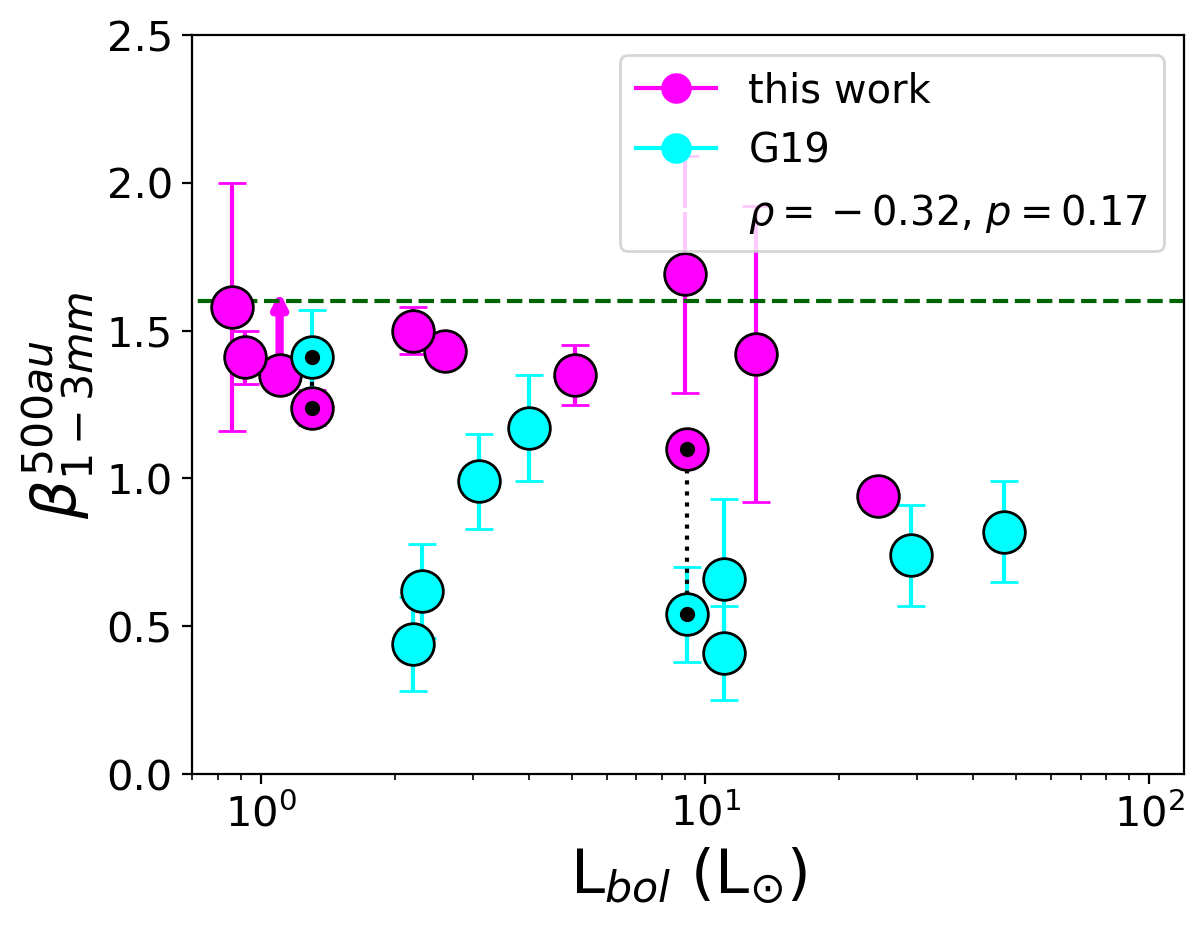}
    \caption{The value of the dust opacity spectral index $\beta$ at 500 au scales from Tab. \ref{tab:all_betas}, as a function of bolometric luminosity of the sources. A black dotted line connects the measured values for IRAS4A and L1527 IRS as measured by G19 and in this work, overlapping within the FAUST and CALYPSO samples (indicated by points with a black central dot). A green dashed line represents typical ISM-like values. For IRAS4C a lower limit is symbolised with an upward arrow. The Pearson correlation coefficient $\rho$ and associated p-value are reported in the upper box.}
    \label{fig:beta_lbol}
\end{figure}

\subsection{The ALMA FAUST sample spectral indices}

Among the sources observed by FAUST and studied in this work (Tab. \ref{tab:faust_sample}), we find five for which no envelope is detected either at 3.1 mm or in both bands, and thus we cannot derive a spectral index for its emission. For the remaining eight sources, we find mostly flat profiles of $\alpha$ and $\beta$ at envelope scales, with values resembling the ones of the diffuse ISM, and thus not showing evidence for early grain growth. Only in the cases of IRAS4A (this work) and L1527 (pilot study by \citealt{Cacciapuoti2023}) we observe a $\beta$ gradient. These two sources are in overlap with the PdBI CALYPSO sample and thus we can compare our results to the ones of \citet{Galametz2019}. In particular, we find consistent values of $\alpha$ and $\beta$ in the uv-plane for both. 

Moreover, thanks to the overall higher sensitivity and resolution of the FAUST data, we could constrain the spectral index of a few bright sources also in the image plane, providing an intuitive display of its variations on the 2D plane of the sky: this was done for IRAS15398-3359 (Fig. \ref{fig:iras15398_alpha_map}), and BHB07-11, VLA1623A and IRAS4A (Fig. \ref{app:2dmaps}). 

The former shows ISM-like $\beta$ along outflow cavity walls as described in section \ref{sec:cavities}; BHB07-11 and VLA1623A show lower spectral indices towards the central 300 au (contaminated by their disks) and ISM-like values where the contamination of this regions fade ($\gtrsim$ 500 au), consistently with the values recovered in the uv-plane after subtraction of the central disk component (in Fig. \ref{fig:alphas_env}). 
Finally, the spectral index of IRAS4A varies between 2.5 and 3.1 between 50 and 500 au (section \ref{sec:iras4a}). As the inner 200 au of IRAS4A are likely optically thick (see section \ref{sec:iras4a}), this gradient might be shaped by optical depth effects (consistently with \citealt{Li2017}, \citealt{Ko2020}). Even so, at 500 au, where the envelope seems to be optically thin, we find a relatively low $\alpha \sim 3.1$ ($\beta \sim 1.1$), which cannot rule the presence of sub-millimetre grains at those scales. Such a $\beta$, hovering close to the limit usually assumed for grain evolution of 1, suggests that constraining parameters that affect $\beta$ secondarily, like dust composition and porosity, could be important.

\begin{figure}[t]
    \centering
    \includegraphics[width=\linewidth] {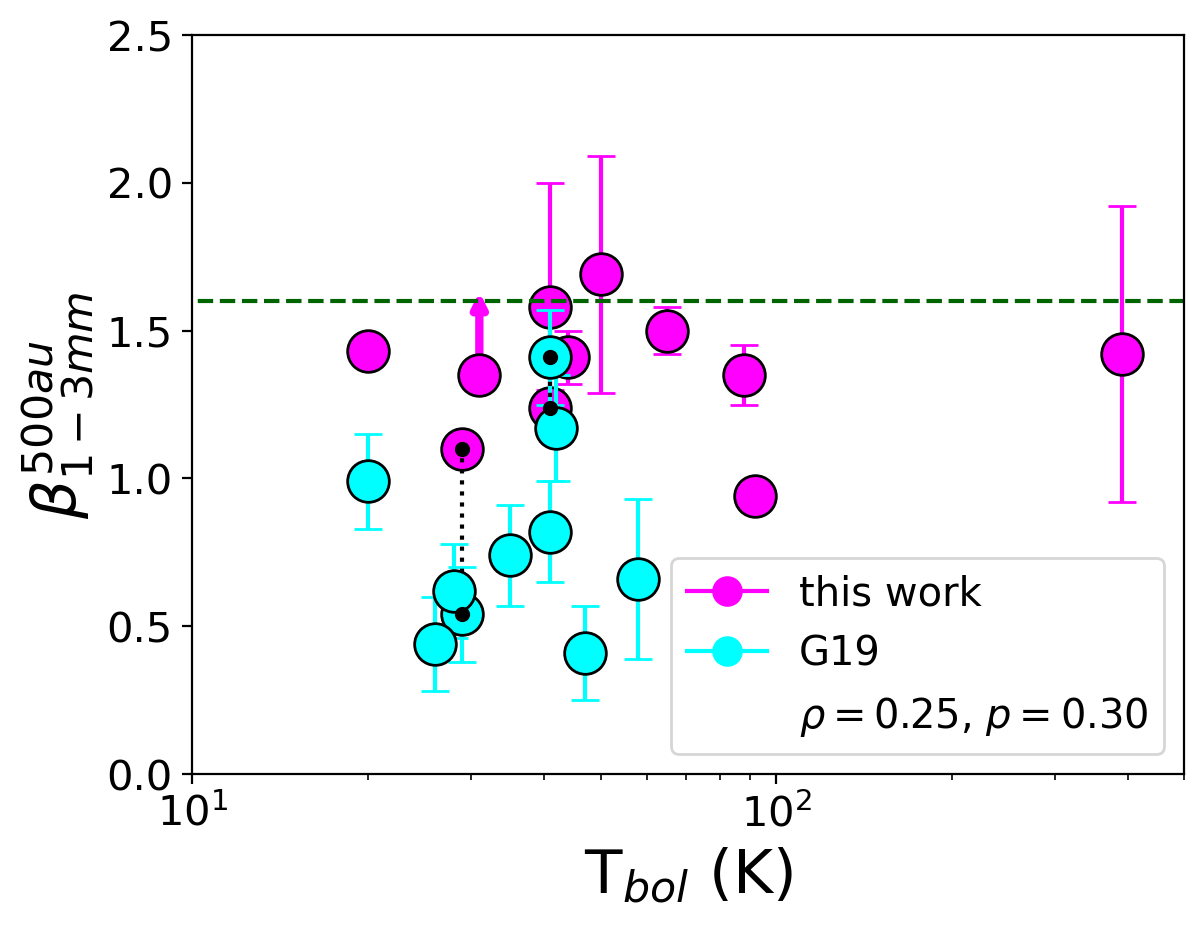}
    \caption{Same as Fig. \ref{fig:beta_lbol}, but for dust opacity spectral index as a function of bolometric temperature of the sources.}
    \label{fig:beta_tbol}
\end{figure}
\subsection{The dust opacity spectral index and protostellar properties}
\label{sec:correlation1}
\begin{figure}[t]
    \centering
    \includegraphics[width=0.95\linewidth] {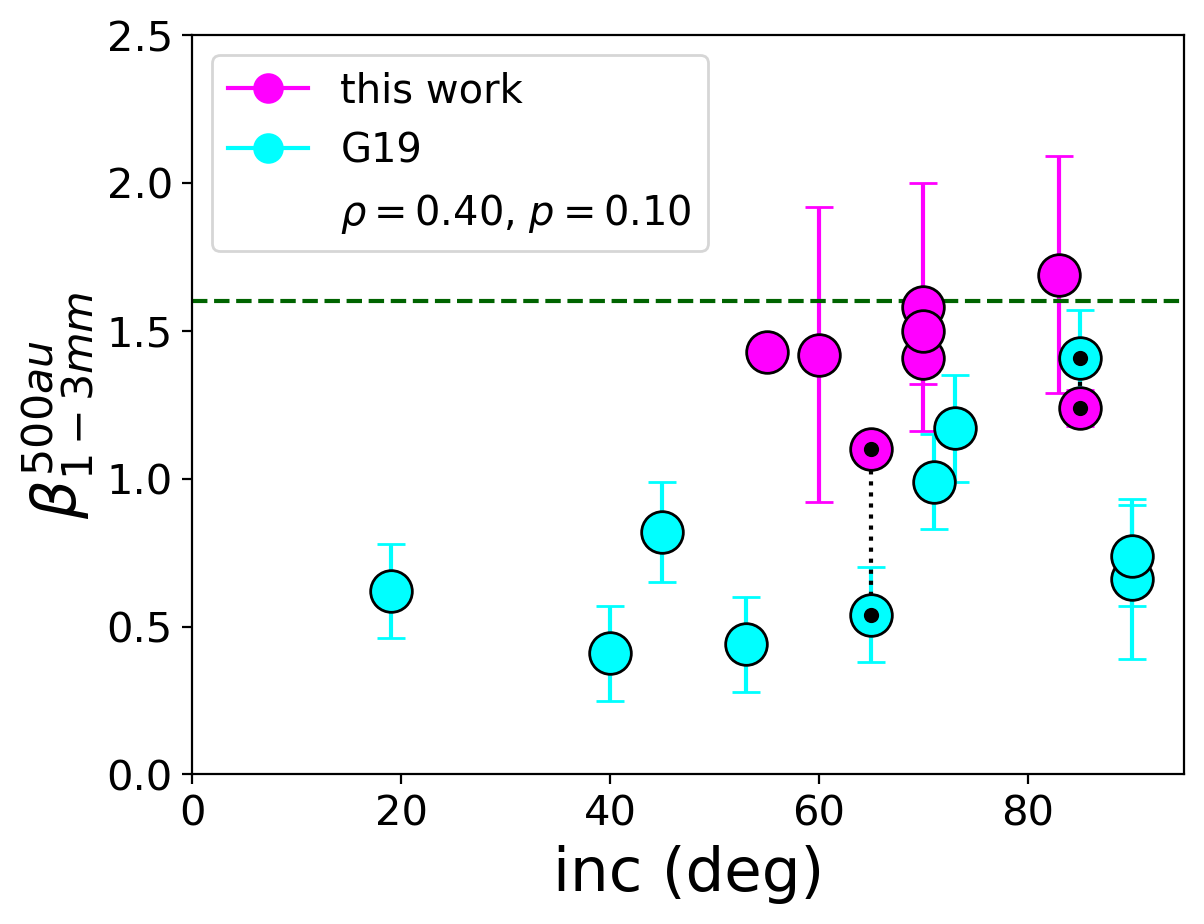}
    \caption{Same as Fig. \ref{fig:beta_lbol}, but for dust opacity spectral index as a function of inclination. Here, an inclination of zero corresponds to a face-on source.}
    \label{fig:beta_inc}
\end{figure}

In order to explore the possibility of correlations between protostellar properties and dust opacity spectral indices in envelopes in a more robust statistical way, we put together the constraints derived in this work (and in the pilot study by \citet{Cacciapuoti2023} for L1527 IRS) with the ones obtained by \citet{Galametz2019} for the CALYPSO sample, in an effort to synthesize all the available robust measurements for envelopes and gain better statistics. Figures \ref{fig:beta_lbol} and \ref{fig:beta_tbol} show the estimated $\beta$ for all objects at a scale of 500 au as a function of protostellar bolometric luminosity L$_{\textrm{bol}}$ and temperature T$_{\textrm{bol}}$. 
No significant correlation is found between any of the pairs. For all correlation tests in this work, we used the \texttt{pearsonr} routine of \texttt{scipy}, which computes the Pearson correlation coefficient $\rho$ and performs a test of the null hypothesis that the distributions underlying the samples are uncorrelated and normally distributed. We report $\rho$ and the p-value of this test in Fig. \ref{fig:beta_lbol} through Fig. \ref{fig:beta_menv}. The reported p-value indicates the probability of an uncorrelated system producing datasets that have a Pearson correlation at least as extreme as the one computed from these datasets (thus, small p-values indicate significant correlations).

It is interesting to note that a correlation with bolometric temperature might have indicated a link between the maximum inferred grain sizes and the evolutionary stage of the object, or in other words, the timescales over which dust might evolve. However, it could also be that the typical timescale of grain growth in envelopes is small compared to the protostellar lifetime and so convergence is reached fast and no difference is seen within the window of 0.1 Myr that we probe. Moreover, using bolometric temperature as an age indicator might only work for very large sample, given the limitations it bears to classify source evolution (e.g., \citealt{Tobin2024}).

A correlation with the internal luminosity might have hinted at changes in $\beta$ due to higher or lower temperatures in the envelope, and thus potential changes in dust composition. 

Finally, to be sure that higher optical depths due to edge-on sources are not confounding the results, we plot $\beta$ against sources inclination. No trend is found (Fig. \ref{fig:beta_inc}).

\subsection{The dust opacity spectral index and protostellar envelope mass}
\label{sec:correlation2}

\begin{figure}[t]
    \centering
    \includegraphics[width=\linewidth] {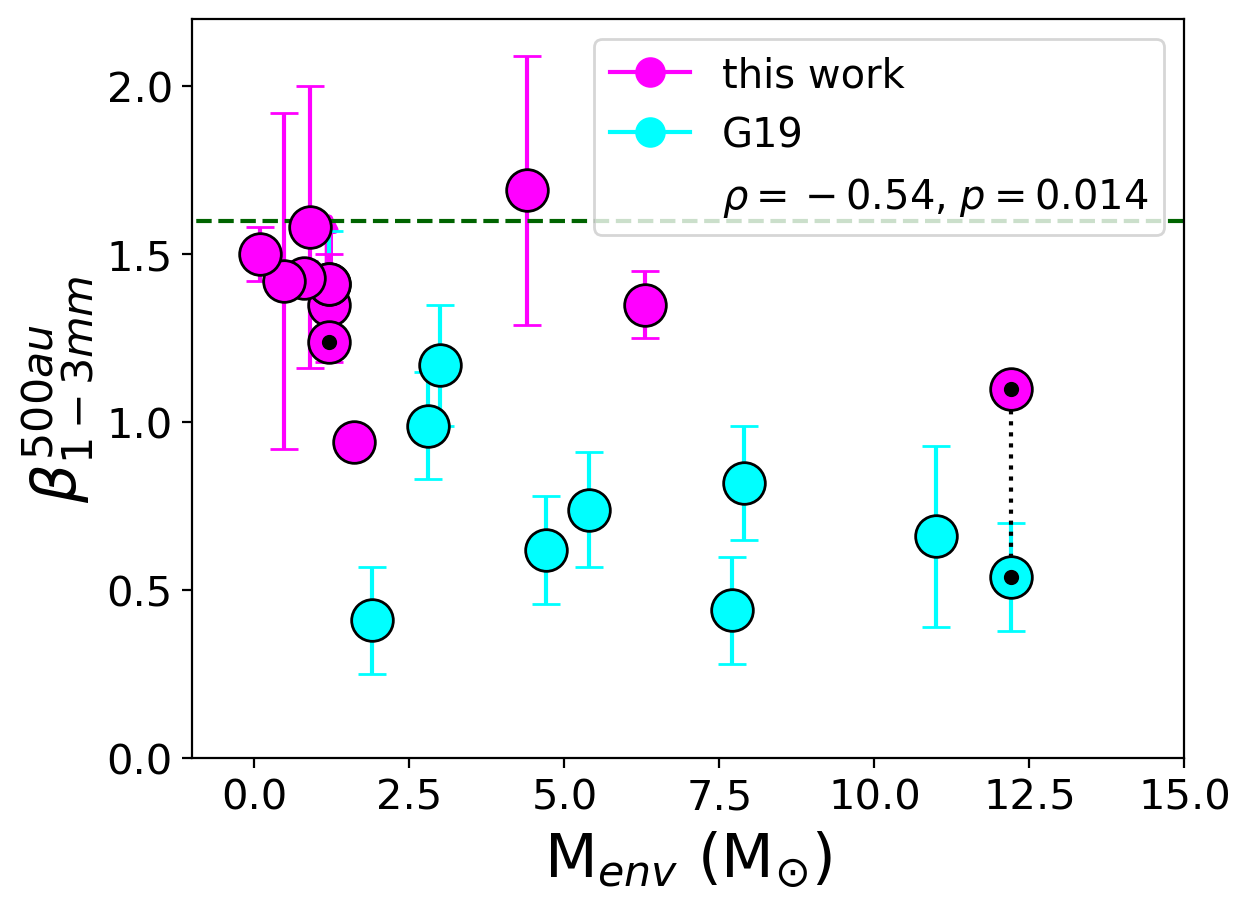}
    \caption{Same as Fig. \ref{fig:beta_lbol}, but for dust opacity spectral index as a function of envelope mass.}
    \label{fig:beta_menv}
\end{figure}

\begin{figure}[t]
    \centering
    \includegraphics[width=\linewidth] {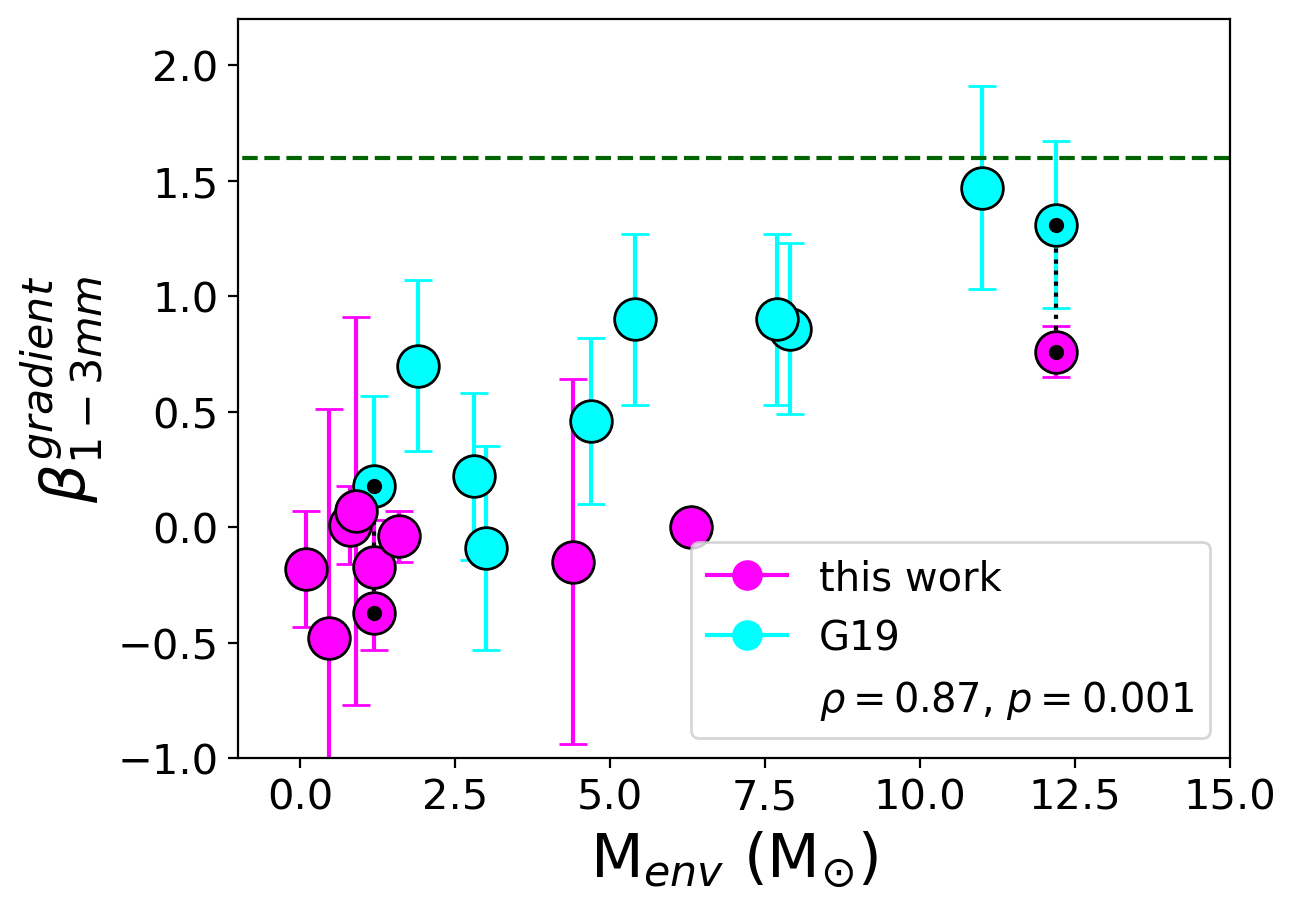}
    \caption{Same as Fig. \ref{fig:beta_lbol}, but for dust opacity spectral index gradient as a function of envelope mass.}
    \label{fig:grad_menv}
\end{figure}

While no link is evident with T$_{\textrm{bol}}$, L$_{\textrm{bol}}$, or source inclination, a significant (see previous section) correlation is found between $\beta$ and the mass of the protostellar envelopes M$_{\textrm{env}}$, as well as between the latter and the gradient of the $\beta$ profile (Fig. \ref{fig:beta_menv} and \ref{fig:grad_menv}). Both correlations were already shown by \citet{Galametz2019}, and here gain in statistical significance. Indeed, while in G19 they both had a p-value of $\sim$0.12, with the new data added in this work the p-values decrease to 0.01 for $\beta$ vs M$_{env}$ and to 0.001 for $\beta^{env}_{grad}$ vs M$_{env}$.
Such correlations have been interpreted as a potential link between the derived $\beta$ and the mass loss rates of outflows by \citet{Cacciapuoti2024}. This scenario was proposed in order to explain the unexpected large grain sizes inferred in the CALYPSO sample envelopes, which seem to be not justified by grain growth models in collapsing environments (e.g. \citealt{Ormel2009}). Indeed, according to \citet{Bontemps1996}, the envelope mass correlates with the CO flux measured in outflows because ejection and accretion are related in a way that stronger accretion will lead to more powerful ejection events. Thus, \citet{Cacciapuoti2024} showed a correlation between the observed CALYPSO $\beta$ values and the mass loss rates associated with their jets and outflows, suggesting it as a tentative evidence of the entraining of large grains from the disk towards the inner envelope. The possibility of such entraining acting at protostellar stages on submillimetre grains is backed up by simulations, e.g., of \citet{Tsukamoto2021} and \citet{Bhandare2024}.
The ALMA FAUST program setups, however, only includes the high excitation SiO ($5-4$) line. A robust identification of the high velocity (HV) and low velocity (LV) ranges of ejection could be tried with such a line but most sources show non-detections (perhaps because of the episodic nature of accretion/ejection during Class 0/I stages), except three: IRAS4A \citep{Chahine2024} , VLA1623A \citep{Codella2024} and IRAS15398-3359 \citep{Okoda2021}. Thus we cannot accurately differentiate between LV outflows and HV jets, as was uniformly done by \citet{Cacciapuoti2024} to measure the related mass loss rates. Still, one could argue that, in light of the correlation found by \citet{Bontemps1996}, almost all FAUST sources (except IRAS4A) should have low mass loss rates and thus that the high $\beta$ values we measured for them would preserve (and strengthen) the correlation of \citet{Cacciapuoti2024}. In order to further explore the transport scenario, new observations are needed to constrain the FAUST sources outflows and/or directly grain properties along outflows across wavelengths (e.g., \citealt{Duchene2023}, \citealt{Sabatini2024}).

Finally, we note that the $\beta$ values measured for the FAUST sources result on average higher than the CALYPSO ones of \citet{Galametz2019}. In the former, the mean $\beta$ is 1.42 $\pm$ 0.06, whereas the average $\beta$ of the latter is 0.78 $\pm$ 0.09. We ran a Kologorov-Smirnov (KS) test to asses the statistical significance of this statement. The KS statistic, measuring a distance metric between the samples, results 0.4 and relative p-value is 0.003. This means that the samples are not consistent with each other at a significance level 3$\sigma$.
This difference could be due to the nature of the samples, since CALYPSO targeted the very brightest Class 0 sources known at the time, while FAUST includes sources with fainter extended emission. In turn, this difference could be explained in two ways (or a combination of them): (i) the more massive envelopes of CALYPSO might be so dense in their central regions that the 1 mm emission is partially optically thick and thus it artificially lowers $\alpha$ and $\beta$, and/or (ii) more massive envelopes are more favorable sites for grain growth due to their higher inner densities. While point (i) seems to be unlikely given the typical densities (n$_{H_2}$ $<$ 10$^8$ cm$^{-3}$) and temperatures (T $\sim$ 20 K) assumed for these environments, further work could more precisely constrain these parameters and uncover the unexpected. However, this degeneracy remains an open problem that can be tackled with radiative transfer post-processing of models of optically thick envelopes with or without grown grains, and multi-wavelength data analysis susceptible to changes in temperature, optical depth and dust emissivity altogether.




\section{Conclusions}
\label{conclusions_faust}

While dust properties have been extensively surveyed in protoplanetary disks and across the ISM at submillimetre wavelengths, a gap remained in what we know about them through the collapsing envelopes that feeds newborn stars and their disks. This gap was largely due to: (i) the difficulty of mapping the thermal emission from envelopes, which is much fainter than the one recovered from their associated disks; (ii) doing so in a multi-scale fashion in order to follow dust evolution across environments that span from $\sim$50 to $\sim$5000 au; and (iii) disentangling the contribution of the high signal-to-noise ratio disks from the envelope in order to measure independent spectral indices. For this reason, only few objects have been studied so far at this early evolutionary stage.

We measured the submillimetre dust spectral index for the ALMA FAUST Large Program sample of class 0/I protostars in order to constrain the dust distribution of 13 protostellar envelopes. We utilise the methods of \citet{Cacciapuoti2023} to disentangle disk and envelope emission in order to accurately derive dust opacity spectral indices of the larger scale emission. In this study, we studied 13 sources in a robust and consistent way and could measure the envelope dust emission spectral index for 11 of them. Adding to previous 9 measurements in the literature (G19), we build a more statistically relevant sample. On the base of this study we draw the following conclusions:

\begin{itemize}
   \item[$\bullet$] The large-scale dust emission of most FAUST sources shows flat spatial profiles ($\beta^{env}_{grad} \sim 0$) with ISM-like values ($\beta \sim 1.4 - 1.8$), which indicate a dust size distribution with maximum grain sizes less than 100 $\mu$m, for eight sources.

   \item[$\bullet$] The dust opacity spectral index of L1527 IRS, L1551-IRS5, and IRAS4A show lower $\beta$ and/or spatial gradients. L1527 IRS is discussed by \citet{Cacciapuoti2023}, and varies from 1 to 1.5 across envelope scales. L1551-IRS displays a flat $\beta \sim 1$. Finally, for IRAS4A, $\beta$ varies from 0.5 to 1.1. These gradients could potentially be attributed to a combination of grain optical properties variation (e.g., dust growth) with the relatively higher optical depth of the innermost regions of these sources. 

   \item[$\bullet$] Compiling the previously known ($n=9$) 1 - 3 mm spectral indices of protostellar envelopes from \citet{Galametz2019} with the ones of this work ($n=11$), we show how the protostellar envelopes spectral index are intermediate between the ones from more evolved protoplanetary disks and the larger-scale ISM, suggesting a bridge between environments in terms of dust evolution (Fig. \ref{fig:alpha_disks_envs}). 
   
   \item[$\bullet$] We find no correlation of $\beta$ with source evolutionary stage, as tentatively identified by the bolometric temperature T$_{\textrm{bol}}$, nor with the internal source luminosity L$_{\textrm{bol}}$, or source inclination. 

   \item[$\bullet$] We statistically strengthen the correlation already suggested by \citet{Galametz2019} between $\beta$ (or its gradient) and envelope mass M$_{\textrm{env}}$. The fact that the inferred dust opacity spectral index lowers in more massive envelopes could potentially be tracing high optical depths at 1 mm of the innermost regions of the more massive protostellar envelopes, or suggesting that more massive envelopes are favorable sites for grain growth (or a combination of both). 

   \item[$\bullet$] The extended continuum emission of IRAS15398-3359, L1551-IRS5, RCrA-IRS7B, and Elias 29 is co-spatial with known outflow cavities. These sources are characterised by a ISM-like spectral index, indicating small dust grains along the cavity walls carved by the outflows, except for L1551-IRS5 (see also \citet{Sabatini2025}). Whether these grains are lifted from the disk or are pristine envelope grains detected at compressed cavity walls remains unclear. 

   \item[$\bullet$] The young protoplanetary disks of the FAUST sample are compact (down to $<$9 au; see also Maureira et al., in prep.) for a follow up at higher resolution) and all consistent with optically thick emission ($\alpha \sim 2.0$), except for the ones of VLA1623A ($\alpha \sim 2.3$), BHB07-11 ($\alpha \sim 2.7$) and IRAS4A ($\alpha \sim 2.6$). See Table \ref{tab:compact_sources_alpha}.
    
\end{itemize}

Pushing the limits of ALMA in resolution, recoverable scales, sensitivity, and frequency ranges today allows us to isolate and study the continuum emission of relatively faint, extended envelopes infalling onto young protostars. Increasing the number of sources for which these measurements are performed and refine the methodologies to obtain such estimates is of critical importance to constrain the dust size distribution at the early stages of collapse. Whether dust growth already starts at envelopes scales and if dust grains are lifted from young disks to envelopes via outflows remain open questions that will need large samples, a more thorough understanding of the temperature profiles in the innermost regions of these environments, and multi-wavelength analyses to weigh the effects of optical depth. 

Moreover, beyond the total intensity submillimetre regime, infrared data will significantly contribute to narrow the frame around envelope dust properties, by investigating extinction profiles and ices variations at these high densities, now within reach thanks to the sensitivity and resolution of JWST (e.g., \citealt{McClure2023}, \citealt{Dartois2024}). Second, submillimetre polarimetric data could be used to further constrain dust properties as it is sensitive to intermediate dust sizes in the 10-100 $\mu$m range (e.g., \citealt{Valdivia2019}, \citealt{Guillet2020}). 

\begin{acknowledgements}
This work was partly supported by the Italian Ministero dell’Istruzione, Universit\`{a} e Ricerca through the grant Progetti Premiali 2012-iALMA (CUP C52I13000140001), by the Deutsche Forschungsgemeinschaft (DFG, German Research Foundation) - Ref no. 325594231 FOR 2634/2 TE 1024/2-1, by the DFG Cluster of Excellence Origins (www.origins-cluster.de). This project has received funding from the European Union’s Horizon 2020 research and innovation program under the Marie Sklodowska-Curie grant agreement No 823823 (DUSTBUSTERS) and from the European Research Council (ERC) via the ERC Synergy Grant ECOGAL (grant 855130). This research has received funding from the European Research Council (ERC) under the European Union’s Horizon 2020 research and innovation programme (MagneticYSOS project, Grant Agreement No 679937). D.J.\ is supported by NRC Canada and by an NSERC discovery Grant. We thank the entire ALMA team for their dedication to provide us with the data we used for this work. GS, LP, ClCo and EB acknowledge the INAF Mini-grant 2023 TRIESTE (“TRacing the chemIcal hEritage of our originS: from proTostars to planEts”; PI: G. Sabatini), the INAF Mini-Grant 2022 “Chemical Origins” (PI: L. Podio), the project ASI-Astrobiologia 2023 MIGLIORA (Modeling Chemical Complexity, F83C23000800005), the National Recovery and Resilience Plan (NRRP), Mission 4, Component 2, Investment 1.1, Call for tender No. 104 published on 2.2.2022 by the Italian MUR, funded by the European Union – NextGenerationEU – Project 2022JC2Y93 ``Chemical Origins: linking the fossil composition of the Solar System with the chemistry of protoplanetary disks'' (CUP J53D23001600006 - Grant Assignment Decree No. 962 adopted on 30.06.2023 by the Italian MUR), and the PRIN-MUR 2020 BEYOND-2p (Astrochemistry beyond the Second period elements, Prot. 2020AFB3FX). E.B. acknowledges contribution of the Next Generation EU funds within the National Recovery and Resilience Plan (PNRR), Mission 4 - Education and Research, Component 2 - From Research to Business (M4C2), Investment Line 3.1 - Strengthening and creation of Research Infrastructures, Project IR0000034 – “STILES - Strengthening the Italian Leadership in ELT and SKA.
LL acknowledges the support of UNAM-DGAPA PAPIIT grants IN108324 and IN112820 and CONACYT-CF grant 263356.
H.B.L. is supported by the National Science and Technology Council (NSTC) of Taiwan (Grant Nos. 111-2112-M-110-022-MY3, 113-2112-M-110-022-MY3).
This paper makes use of the following ALMA data: ADS/JAO.ALMA\#2018.1.01205.L, ADS/JAO.ALMA\#2016.1.01203.S. ALMA is a partnership of ESO (representing its member states), NSF (USA) and NINS (Japan), together with NRC (Canada), MOST and ASIAA (Taiwan), and KASI (Republic of Korea), in cooperation with the Republic of Chile. The Joint ALMA Observatory is operated by ESO, AUI/NRAO and NAOJ. LC thanks Dr. Camila de Sá Freitas and Pastel Esperan\c{c}a for their daily support.
\end{acknowledgements}

\bibliographystyle{aa} 
\bibliography{biblio} 

\appendix

\section{Continuum emission images}
\label{app:images}
\begin{figure*}
    \centering
    \includegraphics[width=\linewidth] {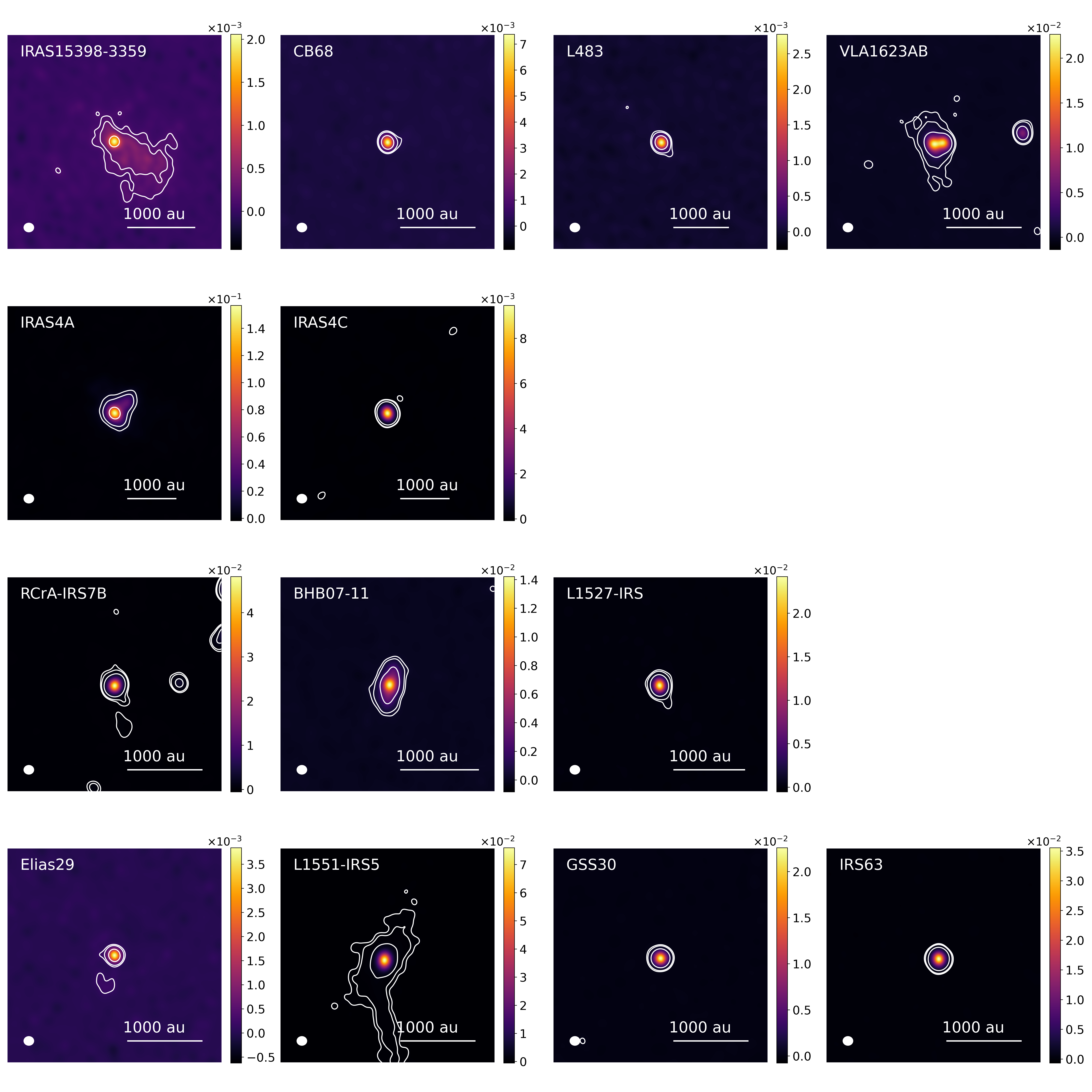}
    \caption{Same as Fig. \ref{fig:sample_1.2 mm} but for the 3.1 mm emission.}
    \label{fig:sample_3.1 mm}
\end{figure*}

While our analysis was mainly conducted in the uv-plane, we here report the continuum emission maps for each source of the FAUST Large Program for completeness at 3.1 mm. These images were also used to visually assess the best model(s) with which to describe the data (see Figures \ref{fig:sample_1.2 mm} and \ref{fig:sample_3.1 mm}). 

The continuum emission around the young protostars in the FAUST sample presents itself as quite diverse.

Some sources (CB68, L483, Elias 29, GSS30, IRS7B, IRS63, IRAS4C) appear very compact in nature, with small to no obvious extended emission in the standard images produced for this work (see previous section). However, the envelope emission becomes clear in the uv-plane, where short-spacing (large-scale) emission is observed in excess of the compact component in some cases (see appendix \ref{app:fits}). 
On the other hand, another class of sources presents much more extended emission in the form of envelopes. BHB07-11 (Fig. \ref{fig:sample_1.2 mm}) displays an inner circumbinary disk and a more extended flattened envelope structure, as already observed by \citet{Alves2017}. A similar structure is observed for VLA1623A, in which a compact central component is associated with an extended envelope around it that reaches beyond a few hundreds astronomical units. see sections \ref{methods_faust}.

The protobinary IRAS4A shows very bright and asymmetric envelopes around both IRAS4A1 (centre) and IRAS4A2 (north-east). see sections \ref{methods_faust} and \ref{sec:iras4a}.

Finally, some of the sources present emission for which the morphology is consistent with that of outflow cavity walls. In one case, for IRAS15398-3359, the extended emission detected at both 1.2 and 3.1 mm is very clearly associated with the cavity walls carved by the molecular outflow of the central protostar, as observed already by \citet{Jorgensen2013} at 0.88 mm. For the compact Elias 29, extended dust emission is also detected southwards and possibly associated with a reported S-shaped outflow (\citealt{Bussmann2007}). see section \ref{sec:cavities}.

\begin{table*}[h!]
\begin{center}
\begin{tabular}{cccc}
\hline \hline
Source & Band & Beam (arcsec x arcsec x deg) & RMS ($\times$ 10$^{-5}$ Jy/beam) \\  \hline 
IRAS15398-3359 & 3& 1.15 $\times$ 1.05 $\times$ 15& 2.2 \\ 
               & 6& 1.15 $\times$ 1.05 $\times$ 15 & 13.0 \\ 
CB68 & 3& 1.15 $\times$ 1.05 $\times$ 15 &  3.6 \\ 
     & 6&1.15 $\times$ 1.05 $\times$ 15 &  5.5  \\ 
L483 & 3& 1.15 $\times$ 1.05 $\times$ 15 & 1.4  \\ 
     & 6& 1.15 $\times$ 1.05 $\times$ 15 & 5.8  \\ 
Elias 29 & 3& 1.15 $\times$ 1.05 $\times$ 15& 2.2 \\ 
         & 6& 1.15 $\times$ 1.05 $\times$ 15 & 5.1 \\ 
VLA1623A & 3& 1.15 $\times$ 1.05 $\times$ 15& 4.3 \\ 
         & 6&1.15 $\times$ 1.05 $\times$ 15 & 35.4 \\ 
GSS30 & 3& 1.15 $\times$ 1.05 $\times$ 15& 3.2 \\ 
      & 6& 1.15 $\times$ 1.05 $\times$ 15 & 100.0 \\ 
RCra IRS7B & 3& 1.15 $\times$ 1.05 $\times$ 15 & 3.5 \\ 
           & 6& 1.15 $\times$ 1.05 $\times$ 15& 64.0 \\ 
BHB07-11   & 3& 1.15 $\times$ 1.05 $\times$ 15 & 4.4 \\ 
           & 6& 1.15 $\times$ 1.05 $\times$ 15 & 36.0 \\ 
IRS63 & 3&  1.15 $\times$ 1.05 $\times$ 15& 2.3 \\
      & 6& 1.15 $\times$ 1.05 $\times$ 15& 20.7 \\ 
NGC1333 IRAS4A & 3&1.15 $\times$ 1.05 $\times$ 15& 16.0 \\ 
               & 6& 1.15 $\times$ 1.05 $\times$ 15 & 390.0  \\ 
NGC1333 IRAS4C & 3& 1.15 $\times$ 1.05 $\times$ 15 & 0.7 \\ 
               & 6& 1.15 $\times$ 1.05 $\times$ 15 & 4.0 \\ 
L1527 IRS & 3& 1.15 $\times$ 1.05 $\times$ 15& 2.6 \\ 
          & 6& 1.15 $\times$ 1.05 $\times$ 15 & 22.5 \\ 
L1551-IRS5 & 3 & 1.15 $\times$ 1.05 $\times$ 15 & 2.3 \\
            & 6 & 1.15 $\times$ 1.05 $\times$ 15& 111.0 \\
\hline

\end{tabular}
\end{center}
\caption{Technical details of the FAUST maps at 1 mm (Fig. \ref{fig:sample_1.2 mm}) and at 3.1 mm (Fig. \ref{fig:sample_3.1 mm}).} 
\label{tab:map_stats}
\end{table*}





\section{Fits results}
\label{app:fits}
\begin{figure*}[h!]
    \centering
    \includegraphics[width=0.85\linewidth] {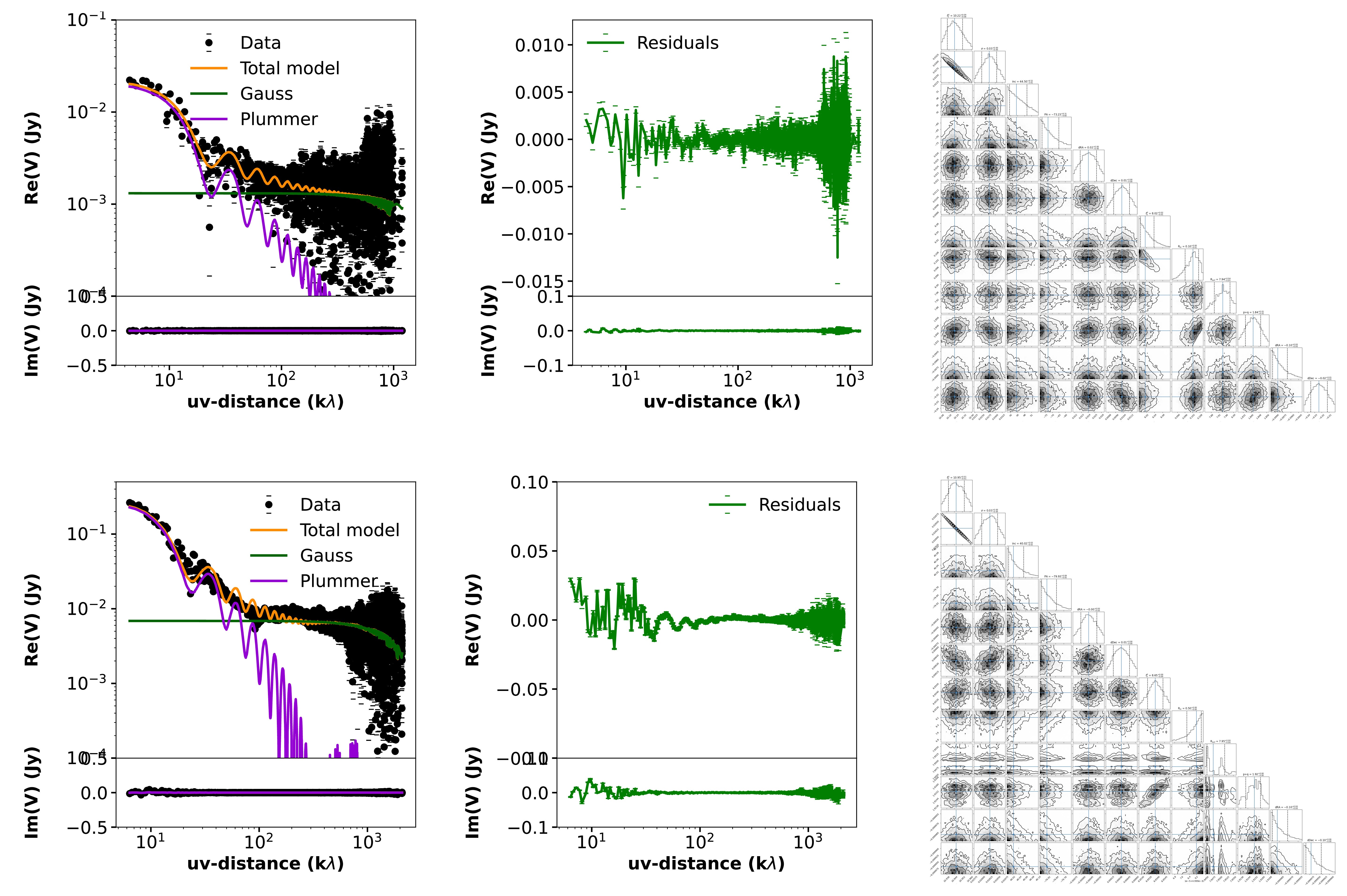}
    \caption{Same as Fig. \ref{fig:cb68_fit}, for IRAS15398-3359.}
    \label{fig:iras153_fit}
\end{figure*}

\begin{figure*}[h!]
    \centering
    \includegraphics[width=0.85\linewidth] {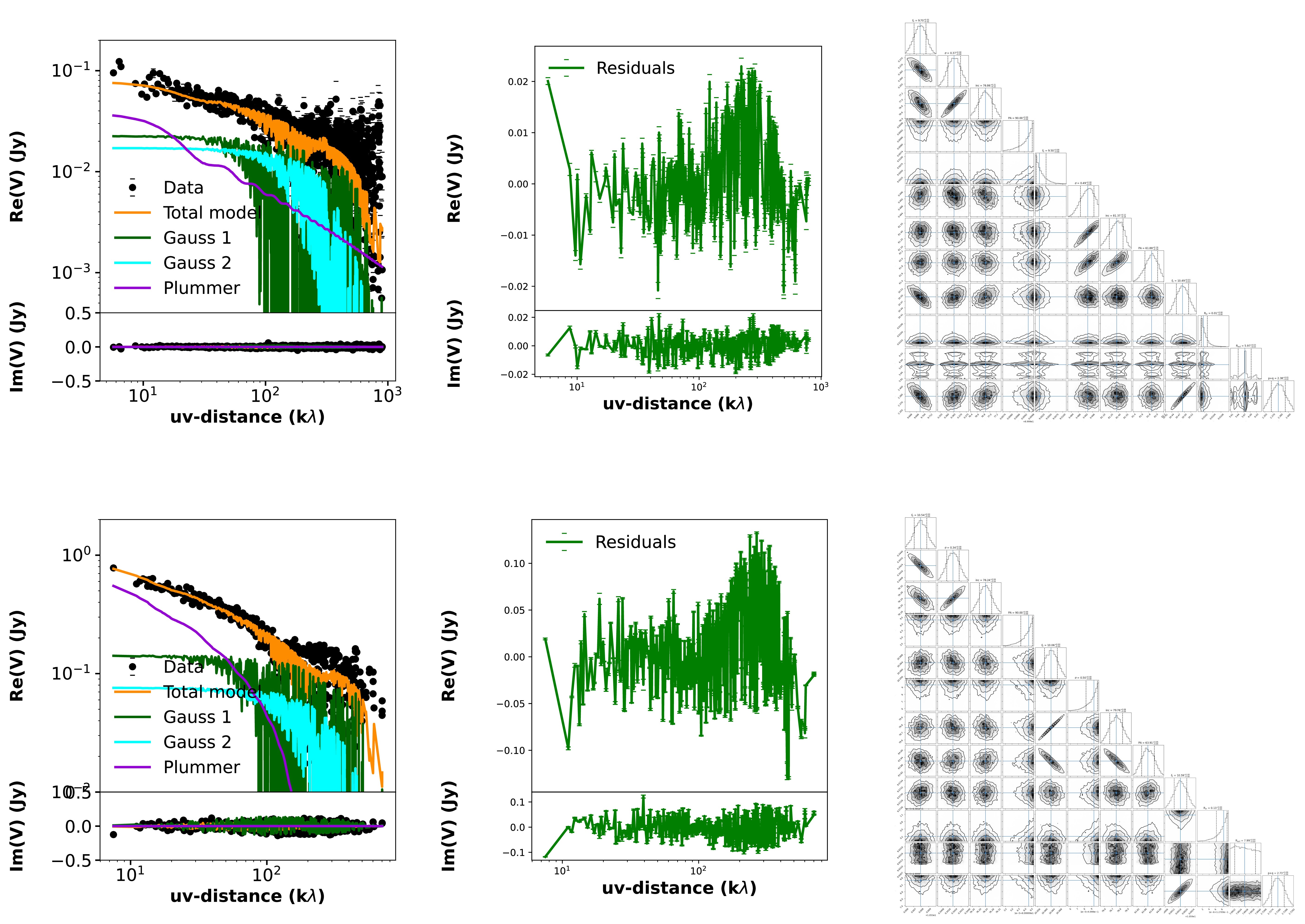}
    \caption{Same as Fig. \ref{fig:cb68_fit}, for VLA1623A, but with two Gaussian components (green and cyan) and one Plummer envelope (violet).}
    \label{fig:vla1623_fit}
\end{figure*}

\begin{figure*}[h!]
    \centering
    \includegraphics[width=0.85\linewidth] {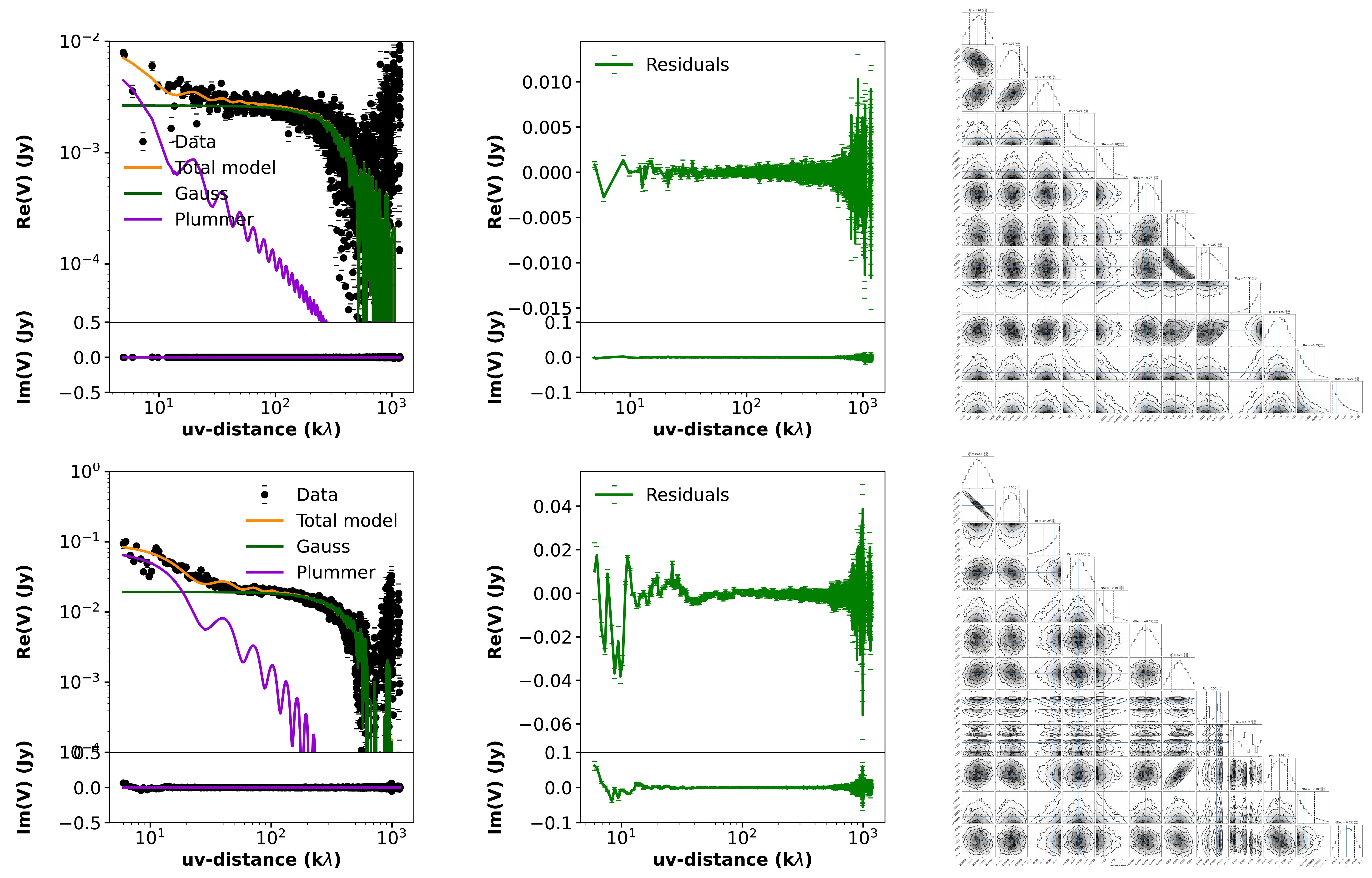}
    \caption{Same as Fig. \ref{fig:cb68_fit}, for L483.}
    \label{fig:l483_fit}
\end{figure*}

\begin{figure*}[h!]
    \centering
    \includegraphics[width=0.85\linewidth] {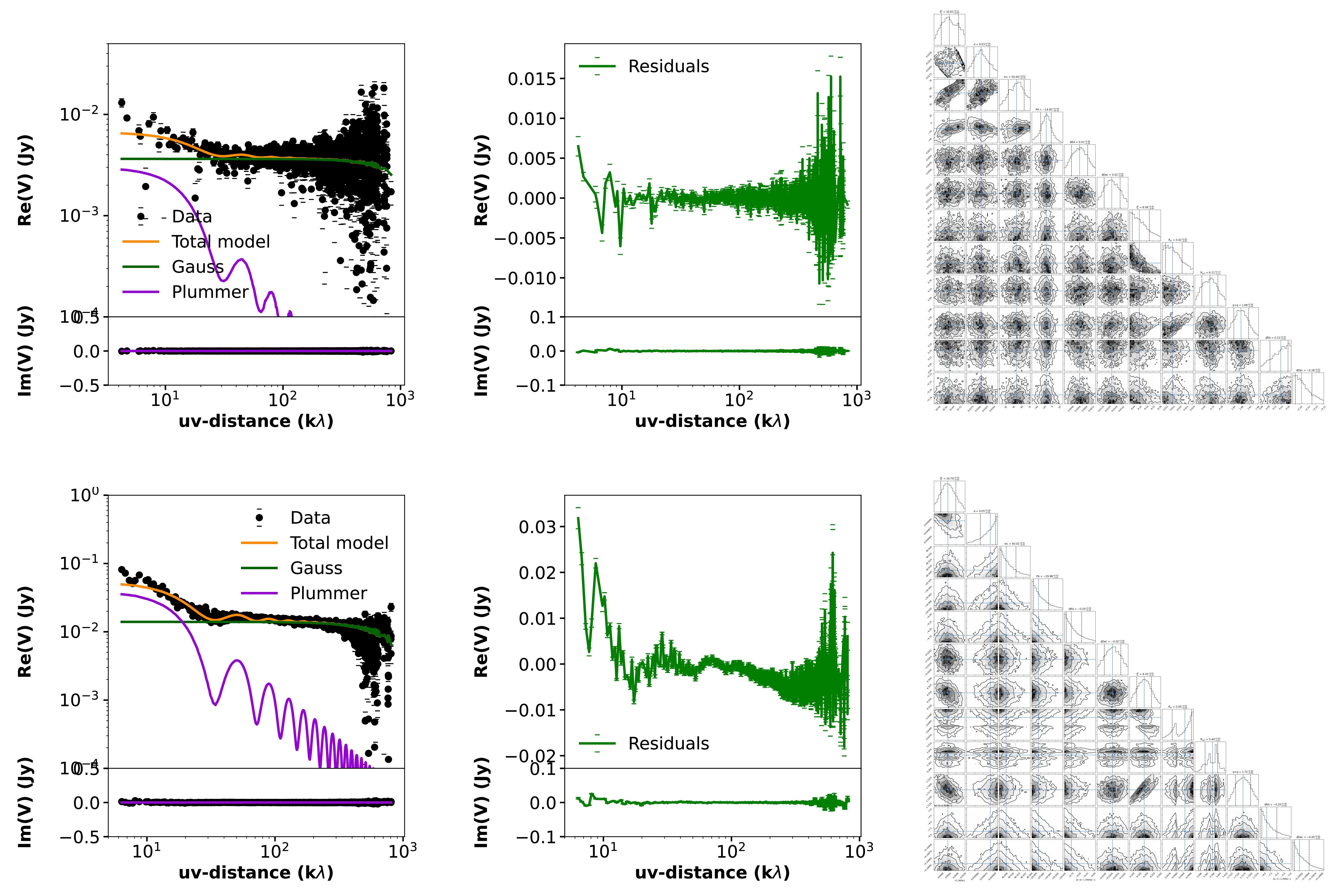}
    \caption{Same as Fig. \ref{fig:cb68_fit}, for Elias 29.}
    \label{fig:elias29_fit}
\end{figure*}

\begin{figure*}[h!]
    \centering
    \includegraphics[width=0.85\linewidth] {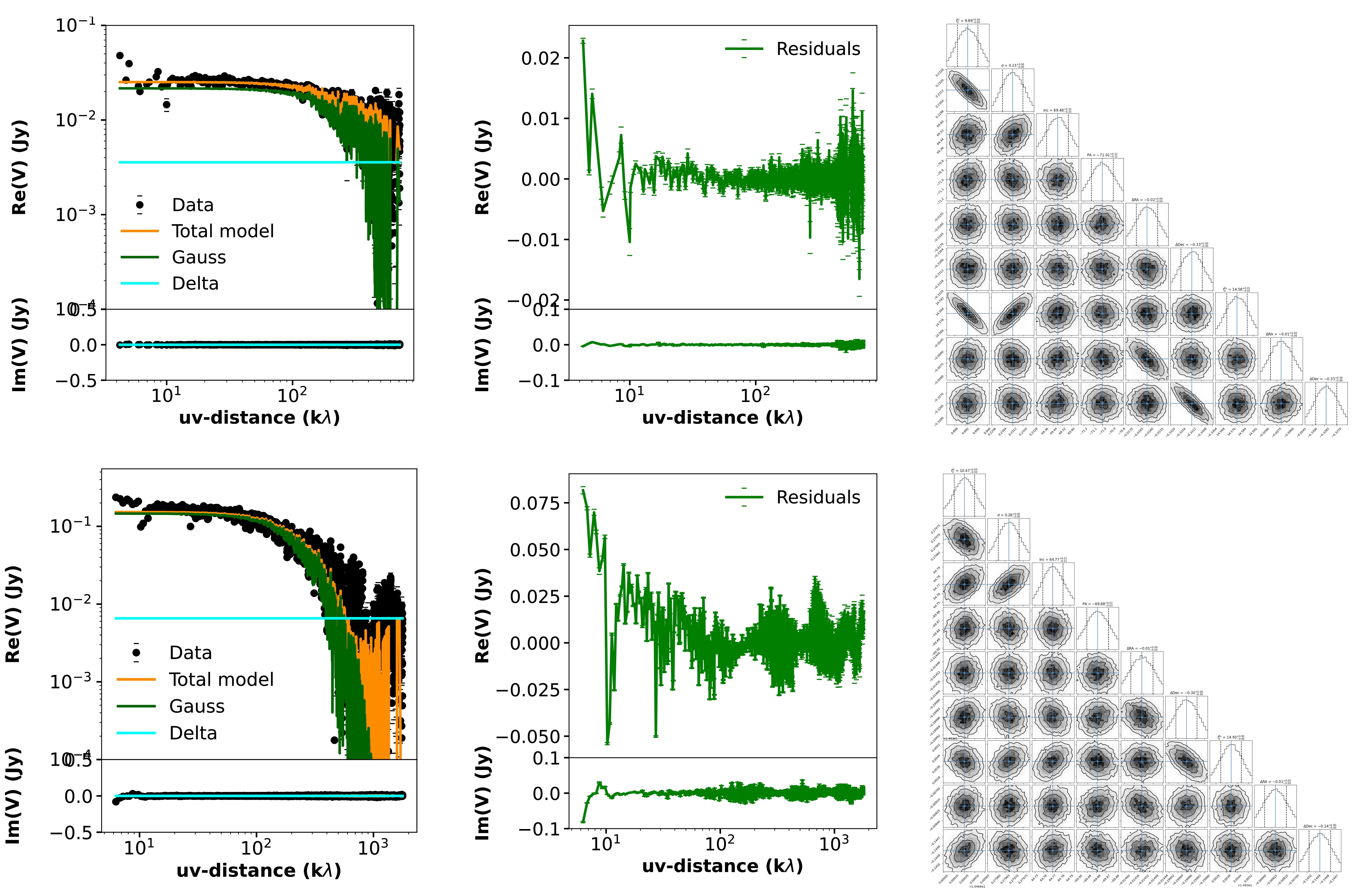}
    \caption{Same as Fig. \ref{fig:cb68_fit}, for GSS30. The 3.1 mm emission is well fit by a Gaussian component. At 1.2 mm an envelope excess was included.}
    \label{fig:gss30_fit}
\end{figure*}

\begin{figure*}[h!]
    \centering
    \includegraphics[width=0.85\linewidth] {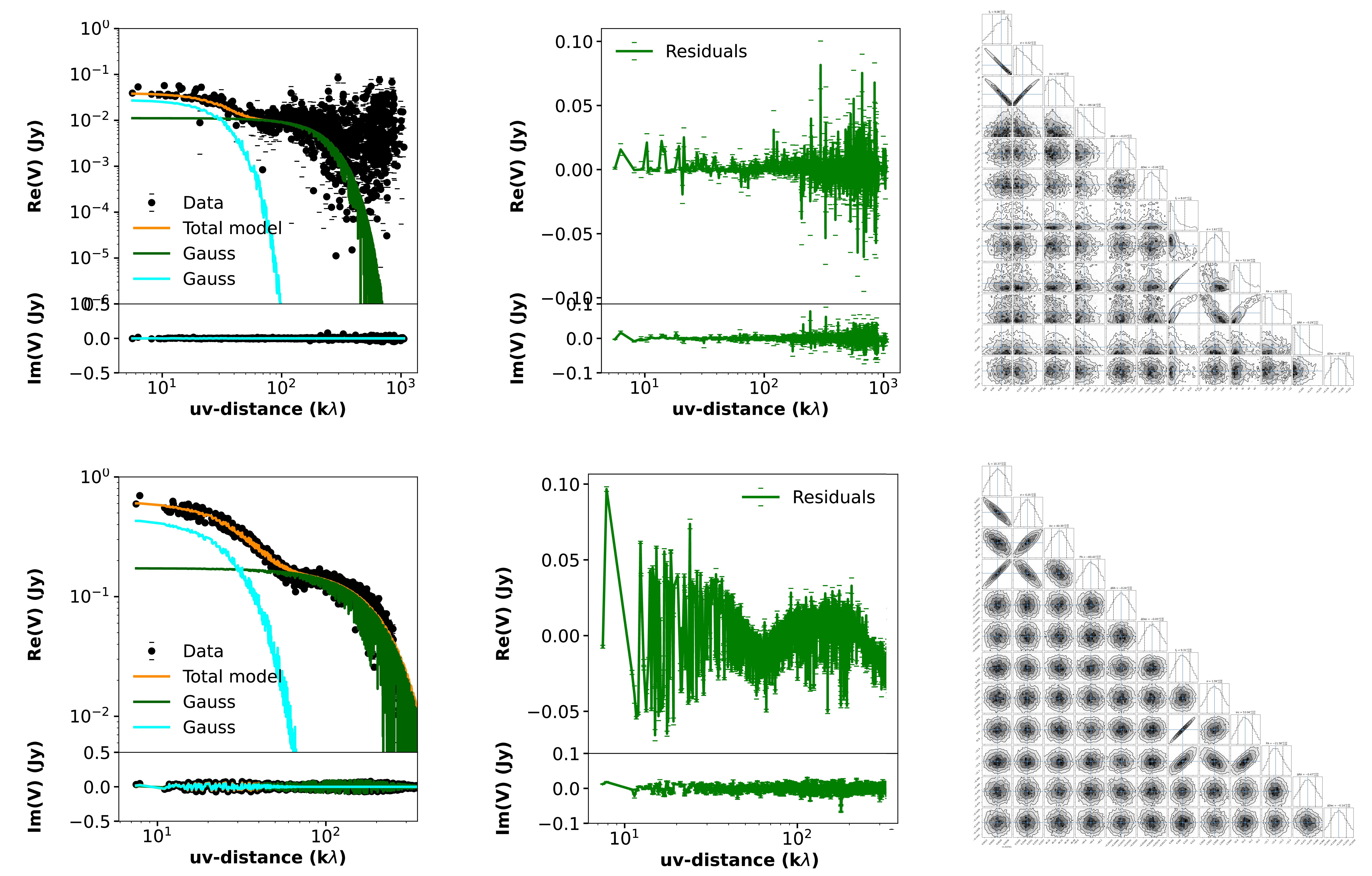}
    \caption{Double Gaussian best fit (orange) is overplotted on the real and imaginary parts of the visibilities for the B3 (upper panel, black points) and B6 (lower panel) observations of BHB07-11.}
    \label{fig:bhb_fit}
\end{figure*}

\begin{figure*}[h!]
    \centering
    \includegraphics[width=0.85\linewidth] {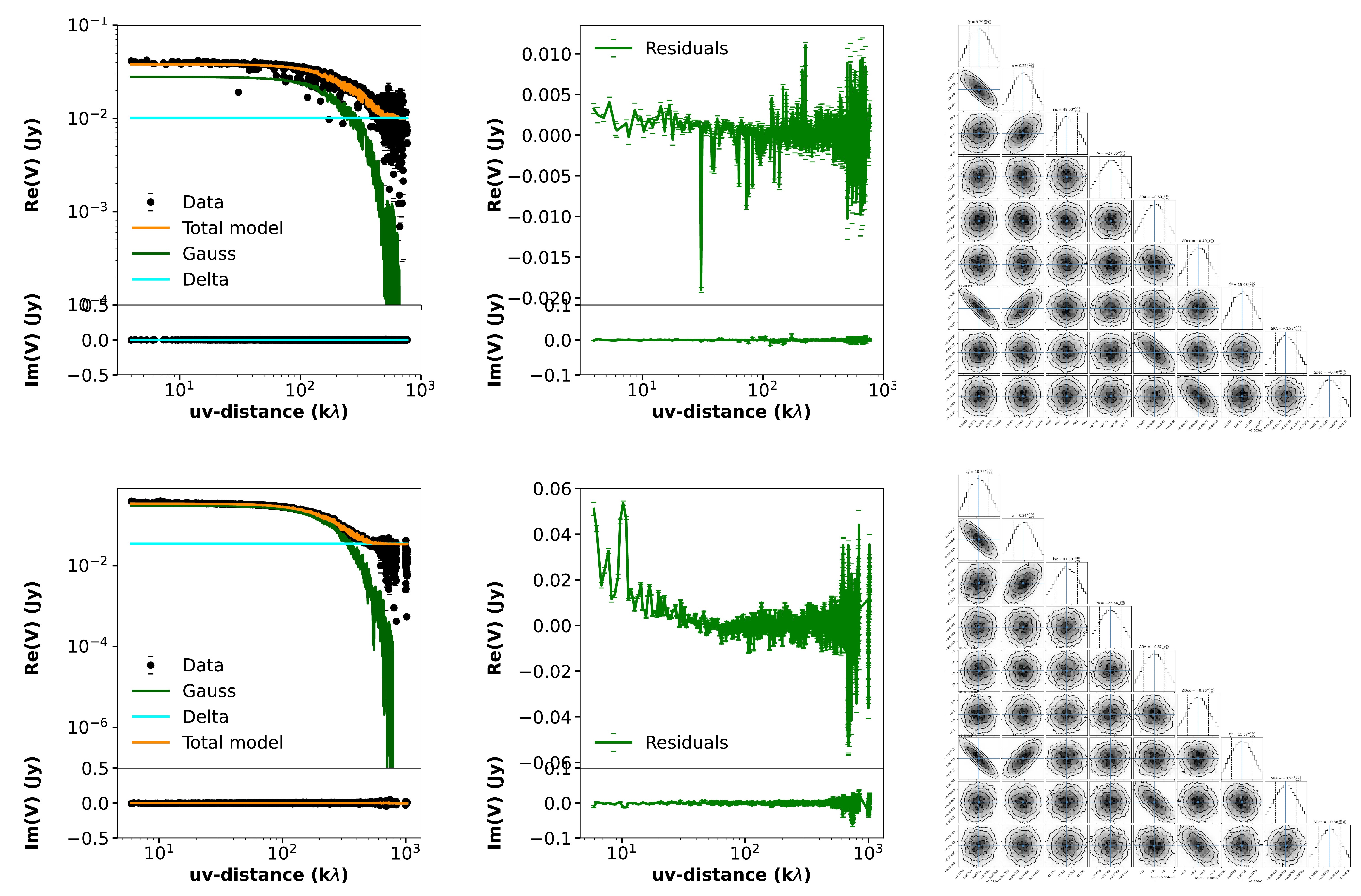}
    \caption{Same as Fig. \ref{fig:cb68_fit}, for IRS63. The 3.1 mm emission is well fit by a Gaussian component and a point-like source.}
    \label{fig:irs63_fit}
\end{figure*}

\begin{figure*}[h!]
    \centering
    \includegraphics[width=0.85\linewidth] {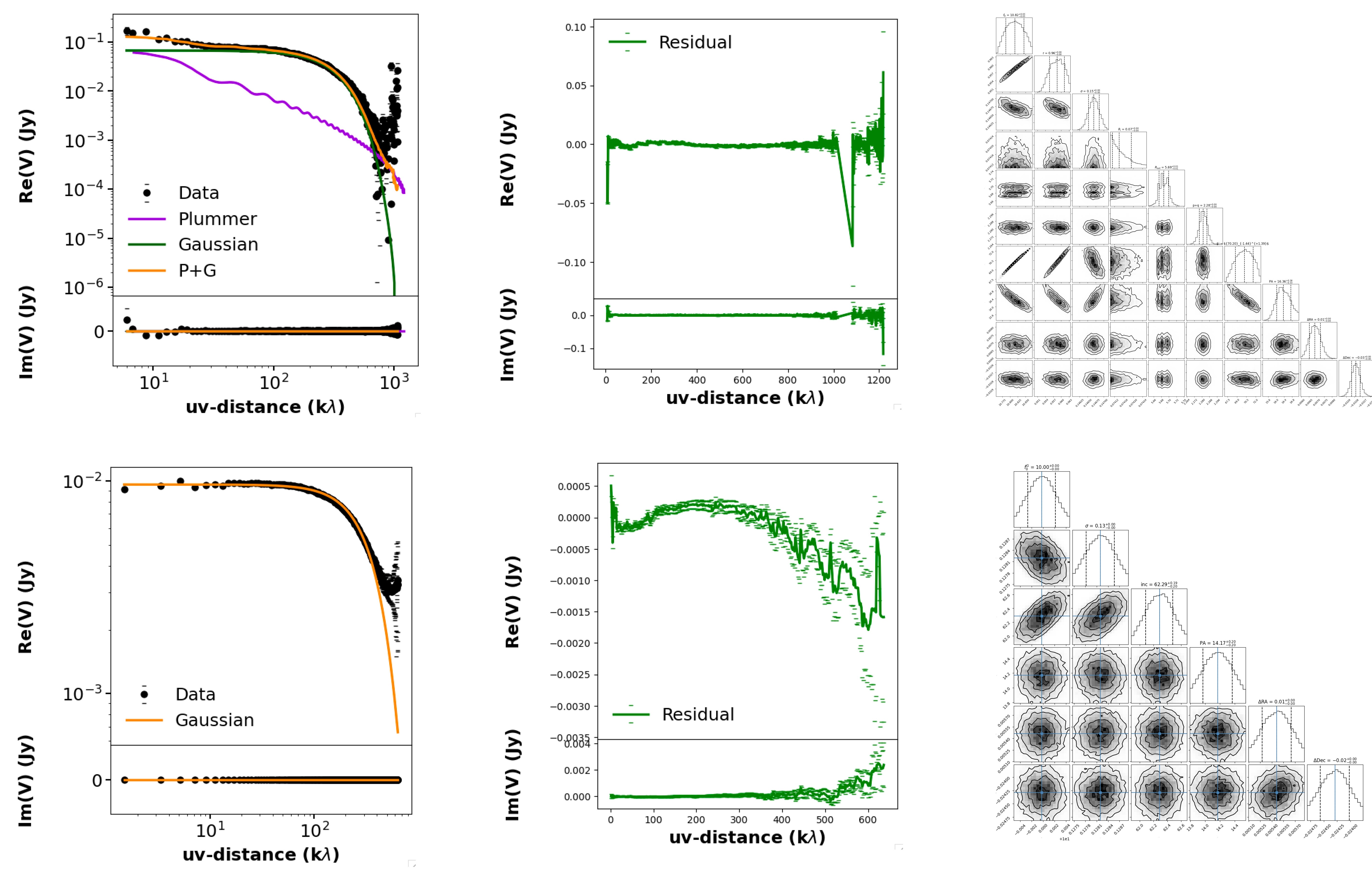}
    \caption{Same as Fig. \ref{fig:cb68_fit}, for IRAS4C. The 3.1 mm emission is well fit by a Gaussian component alone.}
    \label{fig:iras4c_fit}
\end{figure*}

\begin{figure*}[h!]
    \centering
    \includegraphics[width=0.85\linewidth] {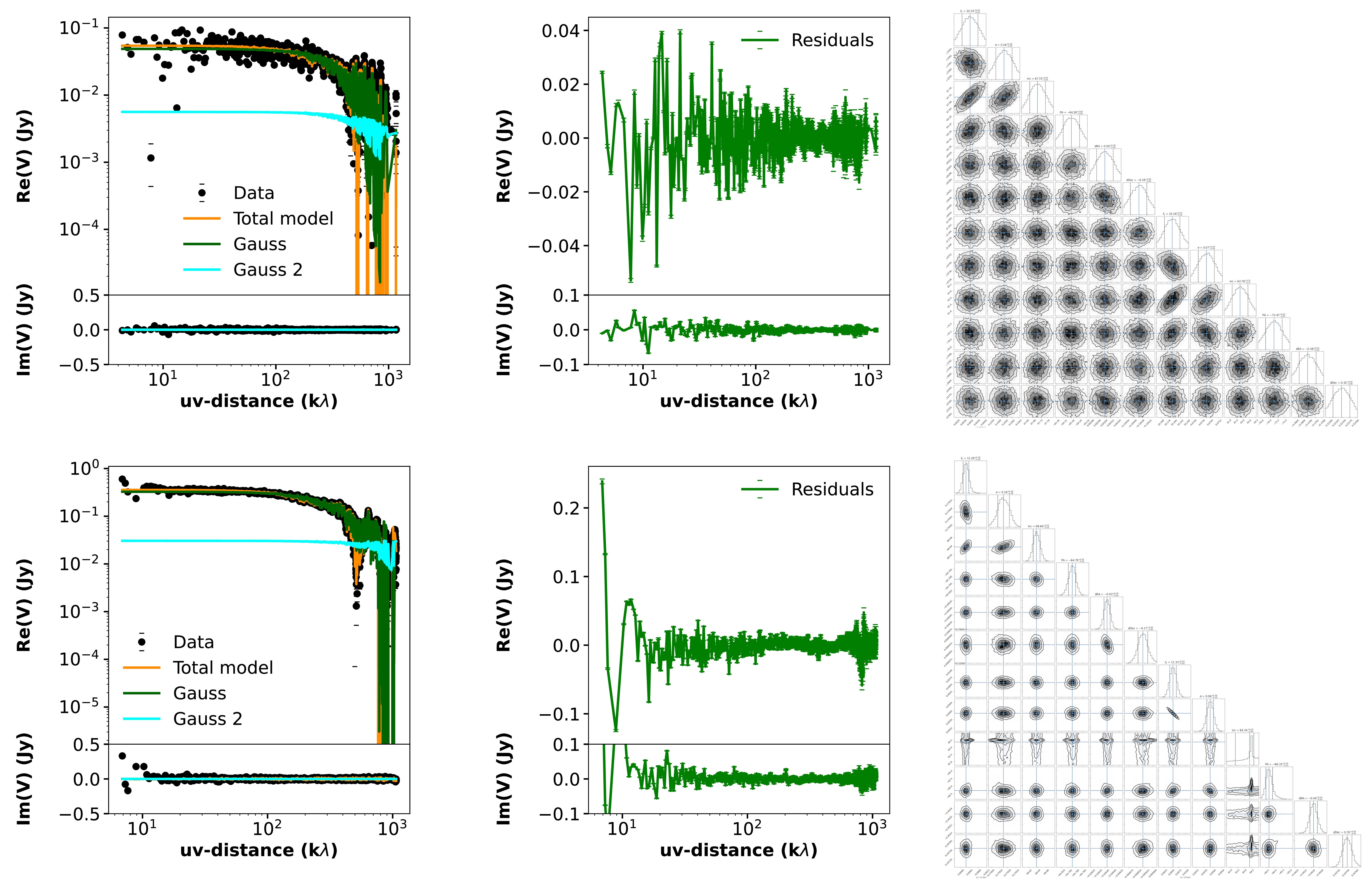}
    \caption{Same as Fig. \ref{fig:cb68_fit}, for IRS7B. The 1.2 mm and 3.1 mm emission is well fit by two compact Gaussians, consistently to what one can appreciate from the sky image in Fig. \ref{fig:sample_1.2 mm}.}
    \label{fig:irs7b_fit}
\end{figure*}

\begin{figure*}[h!]
    \centering
    \includegraphics[width=0.85\linewidth] {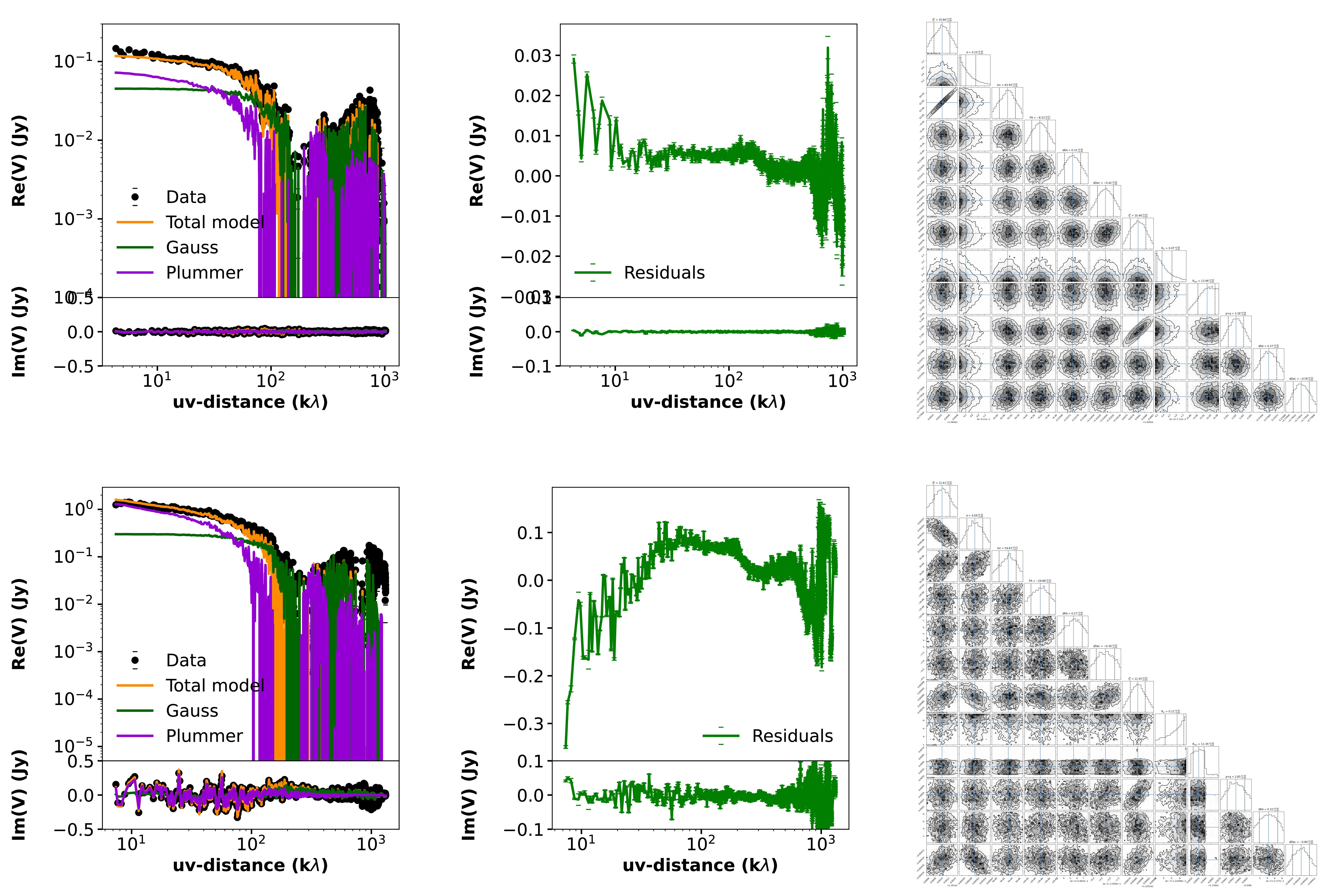}
    \caption{Same as Fig. \ref{fig:cb68_fit}, for L1551-IRS5.}
    \label{fig:l1551_fit}
\end{figure*}

\begin{figure*}[h!]
    \centering
    \includegraphics[width=0.85\linewidth] {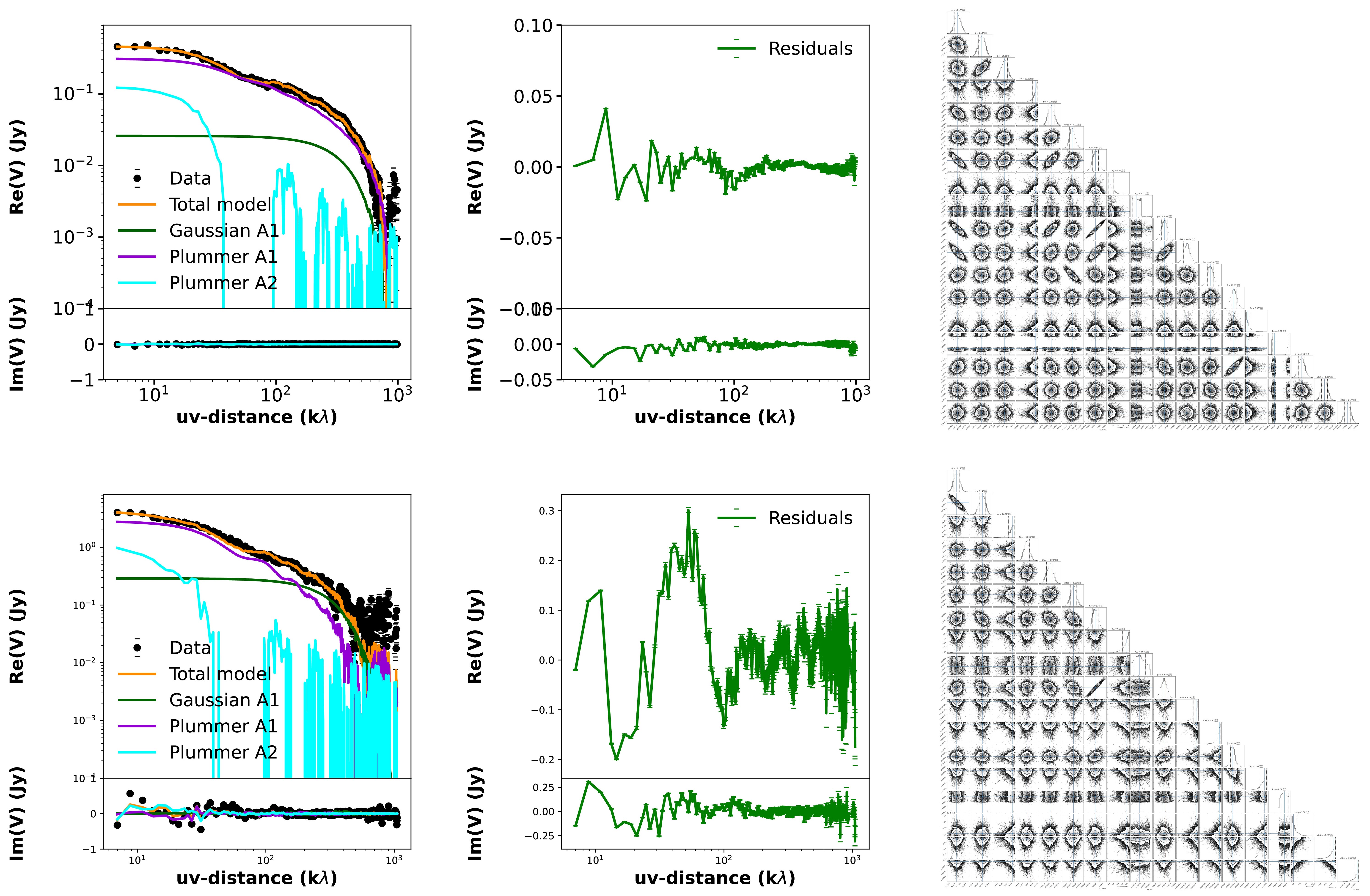}
    \caption{Same as Fig. \ref{fig:cb68_fit}, for IRAS4A. Here a model including a Gaussian (green) and two Plummer profiles (purple and magenta) is adopted to account for the binary nature of IRAS4A, consistently to what one can appreciate from the sky image in Fig. \ref{fig:sample_1.2 mm}.}
    \label{fig:iras4a_fit}
\end{figure*}

Here, we report the best-fit models in the uv-plane along with the residuals, for each of the analysed sources in the same fashion as the example in Fig. \ref{fig:cb68_fit}. The fits are shown on the real and imaginary parts of the visibilities and sampled over the covered (u,v) points, hence the visible wiggles at long baselines. The fits were performed with \texttt{galario} \citep{Tazzari2018}. See Figures \ref{fig:iras153_fit}, \ref{fig:vla1623_fit}, \ref{fig:l483_fit}, \ref{fig:elias29_fit}, \ref{fig:gss30_fit}, \ref{fig:bhb_fit}, \ref{fig:irs63_fit}, \ref{fig:iras4c_fit}, \ref{fig:irs7b_fit}, \ref{fig:l1551_fit}, \ref{fig:iras4a_fit}.



\section{Removal of secondaries}
\label{app:remove_sec}
Here, we show the results of the removal of secondary sources in the field of view of VLA1623A and IRAS4A, as anticipated in Section \ref{methods_faust}. We modeled the secondary sources as gaussians using \texttt{uvmodelfit} CASA routine, converted the output component list using the \texttt{ft} task, input it into the model column of the original data sets and finally subtract the model from the data using \texttt{uvsub}. Figures 

\begin{figure*}[ht!]
    \centering
    \includegraphics[width=\linewidth] {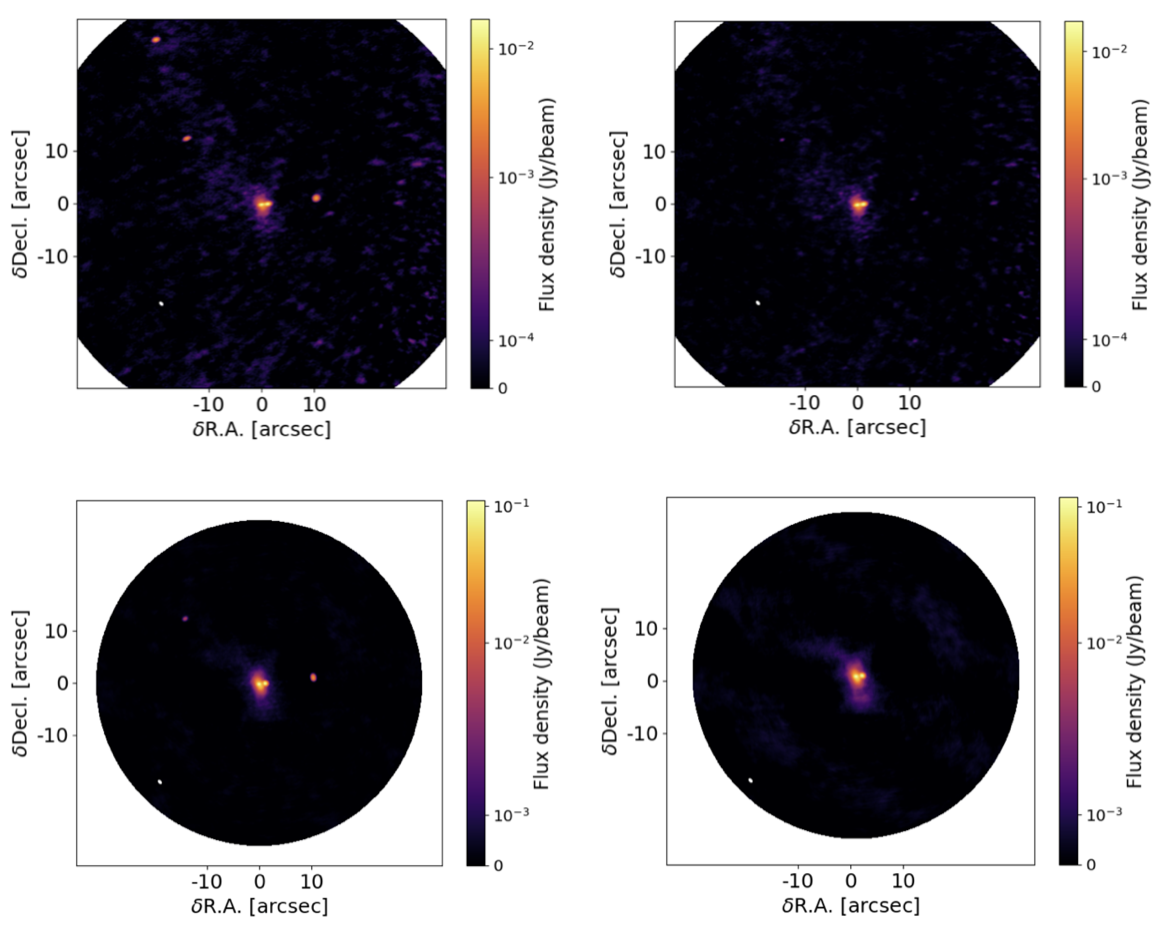}
    \caption{Upper panel: B3 data for VLA1623A before (left) and after subtraction of the three secondary sources in the field of view (right). Lower panel: same as upper panel, but for B6 data and for the only two sources seen in the smaller field of view. The color scale is preserved for fair comparison. The beam is shown as a white ellipse in the lower left corners. The remaining data was modeled with two gaussians for the disks of VLA1623 A and B, and a Plummer envelope centred on VLA1623A. See Fig. \ref{fig:sample_1.2 mm} for a zoom-in.}
    \label{fig:vla1623a_remove}
\end{figure*}

\begin{figure*}[ht!]
    \centering
    \includegraphics[width=\linewidth] {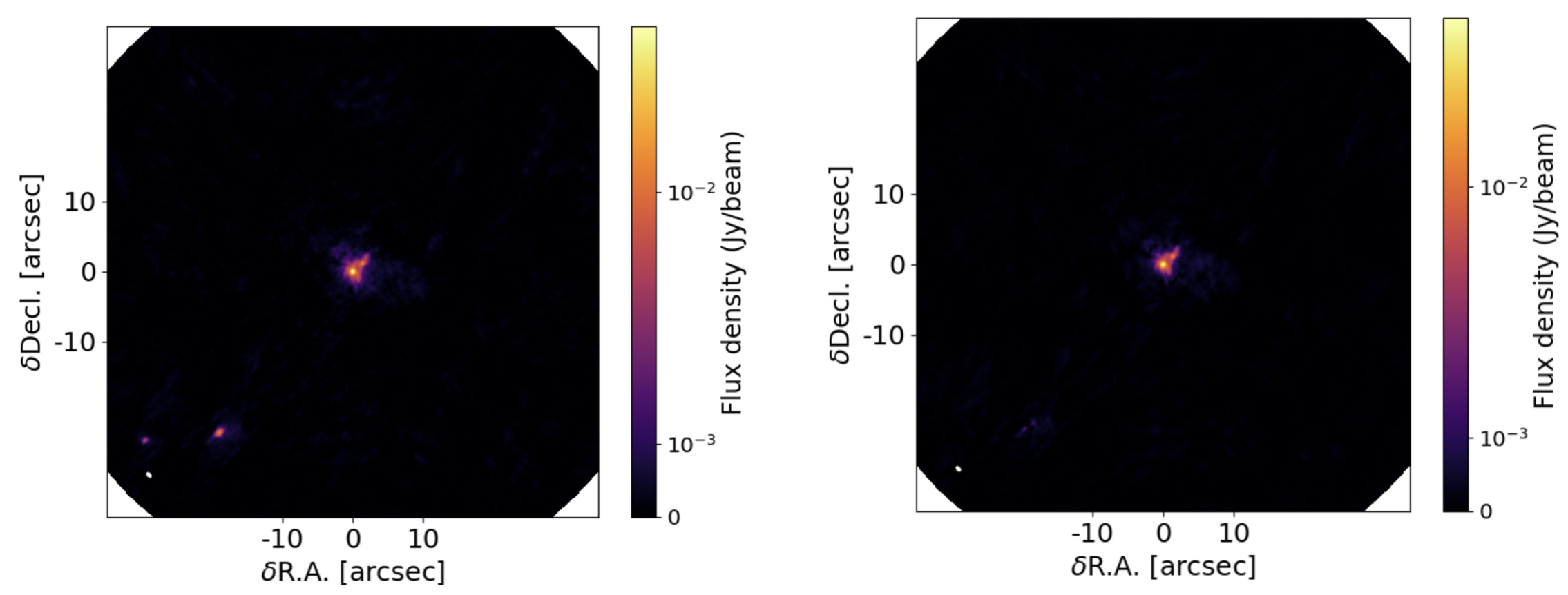}
    \caption{Same as Fig. \ref{fig:vla1623a_remove} but for IRAS4A. Here, the secondary sources in the south-west only appear in the larger B3 field of view, hence the B6 data is not shown. See Fig. \ref{fig:sample_1.2 mm} for a zoom-in.}
    \label{fig:iras4a_remove}
\end{figure*}

\section{2D spectral index maps}
\label{app:2dmaps}
\begin{figure*}
    \centering
    \includegraphics[width=\linewidth] {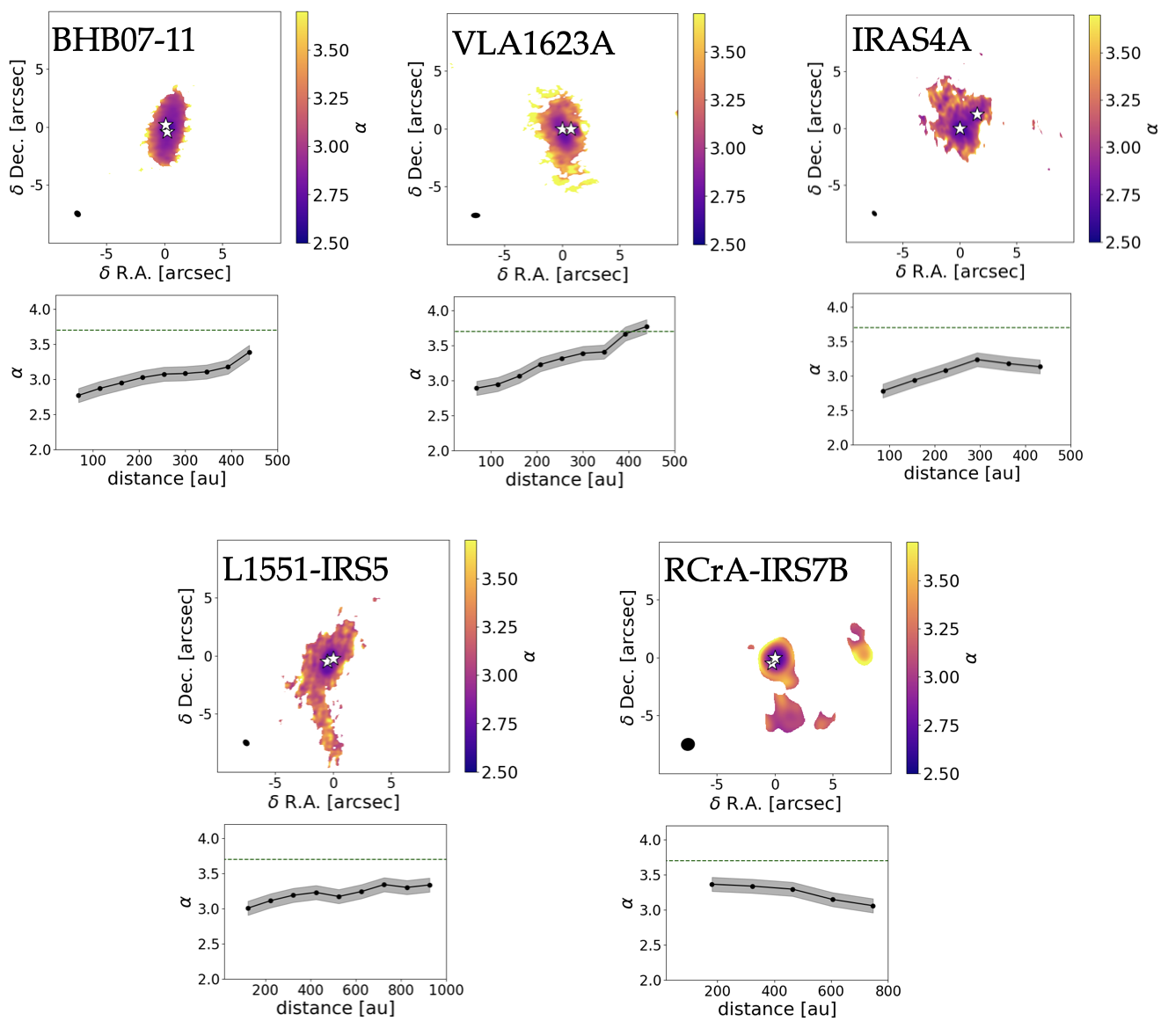}
    \caption{Same as Fig. \ref{fig:iras15398_alpha_map}, but for BHB07-11 (upper left), VLA1623A (upper centre) and IRAS4A (upper right), L1551-IRS5 (lower left), RCrA-IRS7B (lower right). The azimuthally averaged spectral index is sampled on a beam-based cadence (Tab. \ref{tab:2d_alpha_tab}).
    White stars mark source positions. The profile is extracted from concentric annuli around the main target (placed at the centre).}
    \label{fig:2d_maps}
\end{figure*}

As described in section \ref{sec:beta_class0}, we here report in Fig. \ref{fig:2d_maps} the 2D spectral index maps of three of the four brightest envelopes in our sample, for which such an analysis was attempted. The other one, around IRAS15398-3359, is already reported and discussed in section \ref{sec:cavities}. The statistics of the 1.2 and 3.1 mm maps used to generate the 2D spectral index maps are reported in Table \ref{tab:2d_alpha_tab}.

\begin{table*}[h!]
\begin{center}
\begin{tabular}{cccc}
\hline \hline
Source & Band & Beam (arcsec) & RMS ($\times$ 10$^{-5}$ Jy/beam) \\  \hline 
IRAS15398-3359 & 3& 0.46 $\times$ 0.29 & 1.7 \\ 
               & 6& 0.46 $\times$ 0.29 & 1.5 \\  
VLA1623A & 3& 0.73 $\times$ 0.40 & 4.8 \\ 
         & 6& 0.73$\times$ 0.40 & 10.0 \\ 
BHB07-11   & 3& 0.57 $\times$ 0.45 & 2.6 \\ 
           & 6& 0.57 $\times$ 0.45 & 10.5 \\ 
NGC1333 IRAS4A & 3&  0.48 $\times$ 0.31 & 14.0 \\ 
               & 6&  0.48 $\times$ 0.31 & 42.0 \\ 
L1551-IRS5 & 3& 0.43 $\times$ 0.41  & 1.1 \\ 
               & 6&  0.43 $\times$ 0.41  & 26 \\ 
RCrA-IRS7B & 3 & 1.15 $\times$ 1.05 & 3.5  \\
           & 6 & 1.15 $\times$ 1.05 & 64  \\
\hline
\end{tabular}
\end{center}
\caption{Technical details of the FAUST maps at 1.2 mm and at 3.1 mm used to compute the high resolution spectral index maps of Fig. \ref{fig:2d_maps}.} 
\label{tab:2d_alpha_tab}
\end{table*}

\end{document}